\documentclass{crc-book-new}
\usepackage{amsmath, amssymb}
\usepackage{comment}
\usepackage{booktabs}

\usepackage{longtable}
\usepackage{wrapfig}
\usepackage{ifthen}

\usepackage[margin=1cm]{caption}

\usepackage{graphics, graphicx, color, xcolor}

\usepackage[mathscr]{eucal}

\usepackage{pdflscape}
\usepackage{geometry,multicol}  %[centering, margin=20mm]
\newdimen\stockheight
\newdimen\stockwidth
\stockwidth\paperwidth
\usepackage[center,cam]{crop}

\JNMPnumberwithin{equation}{section}
\DeclareMathAlphabet\mathbfcal{OMS}{cmsy}{b}{n}

\newcommand{\mc}[1]{\mathcal{#1}}
\newcommand{\atdm}[1]{\text{\textbf{\~#1}}}
\newcommand{\PDeltaE}{P$\mathrm{\Delta}$E}
\newcommand{\PDeltaEs}{P$\mathrm{\Delta}$Es}
\newcommand{\dsp}{\displaystyle}

\newcommand{\nG}{\mathbb{G}}
\newcommand{\mG}{\mathscr{G}}
\newcommand{\contG}{{\bf{G}}}
\newcommand{\fnum}[1]{ \ifthenelse{\equal{#1}{x}}{f}
                      {\ifthenelse{\equal{#1}{y}}{g}
                      {\ifthenelse{\equal{#1}{z}}{h}{m}}} }
\newcommand{\fden}[1]{ \ifthenelse{\equal{#1}{x}}{F}
                      {\ifthenelse{\equal{#1}{y}}{G}
                      {\ifthenelse{\equal{#1}{z}}{H}{M}}} }
\newcommand{\rgt}{{\footnotesize {\rm Right}}}
\newcommand{\lft}{{\footnotesize {\rm Left}}}

\renewcommand{\evenhead}{Terry J. Bridgman, Willy Hereman}
\renewcommand{\oddhead}{Lax pairs for edge-constrained systems of \PDeltaEs}

\begin{document}

\thispagestyle{empty}

% fixed title in lower cases 09/16/2019
% \Name{Lax Pairs for Edge-constrained Boussinesq Systems of Partial Difference Equations}
\Name{Lax pairs for edge-constrained Boussinesq systems of partial difference equations}

% CRC authors and addresses
\Author{Terry J.\ Bridgman and Willy Hereman}
\Address{Department of Applied Mathematics and Statistics, \\
         Colorado School of Mines, Golden, CO. \\
         e-mail: tbridgma@mines.edu and whereman@mines.edu.}

% ABSTRACT
\begin{abstract}
 The method due to Nijhoff and Bobenko \& Suris to derive Lax pairs for partial
 difference equations (\PDeltaEs) is applied to edge constrained Boussinesq systems.
 These systems are defined on a quadrilateral.
 They are consistent around the cube but they  contain equations defined on the edges of the quadrilateral.

 By properly incorporating the edge equations into the algorithm, it is
 straightforward to derive Lax matrices of minimal size.  The $3$ by $3$ Lax
 matrices thus obtained are not unique but shown to be gauge-equivalent.
 The gauge matrices connecting the various Lax matrices are presented.
 It is also shown that each of the Boussinesq systems admits a $4$ by $4$ Lax
 matrix.  For each system, the gauge-like transformations between Lax matrices
 of different sizes are explicitly given. To illustrate the analogy between
 continuous and lattice systems, the concept of gauge-equivalence of Lax pairs
 of nonlinear partial differential equations is briefly discussed.

 The method to find Lax pairs of \PDeltaEs\ is algorithmic and is being
 implemented in {\sc Mathematica}.
 The Lax pair computations for this chapter helped further improve and extend the capabilities
 of the software under development.
\end{abstract}
%
% INTRO
\section{Introduction}
\label{sec:Introduction}
As discussed by Hietarinta \cite{ref:Hietarinta2018} in volume 1 of this book series,
Nijhoff and Capel \cite{ref:NijhoffCapel1995},
and Bridgman \cite{ref:Bridgman2018}, nonlinear partial ``discrete or lattice" equations (\PDeltaEs),
% fixed contexts 09/16/2019
arise in various contexts.
They appeared early on in papers by Hirota \cite{ref:HirotaI}
covering a soliton preserving discretization of the direct (bilinear) method for nonlinear PDEs
(see, e.g., \cite{ref:Hirota2004}).
In addition, Miura \cite{ref:Miura1968} and Wahlquist \& Estabrook \cite{ref:Wahlquist1973} indirectly contributed
to the development of the theory of \PDeltaEs\ through their work on B{\"a}cklund transformations.
A major contribution to the study of \PDeltaEs\ came from Nijhoff and colleagues
\cite{ref:Nijhoff1991,ref:Nijhoff1983,ref:Nijhoff1982}.
Under the supervision of Capel, the Dutch research group used a direct linearization method and B{\"a}cklund transformations,
in connection with a discretization of the plane wave factor, to derive several \PDeltaEs.
For a detailed discussion of these methods as well as the seminal classification of scalar
\PDeltaEs\ by Adler {\em et al.} \cite{ref:Adler2003,ref:BobenkoSuris2002} we refer to recent books
on the subject
\cite{ref:BobenkoSuris2008,ref:HietarintaJoshiNijhoff2016,ref:Hydon2014,ref:Levietal2011,ref:Levietal2017}.

\begin{figure}[h]
 \centering
\centerline{
\includegraphics[width=120mm]{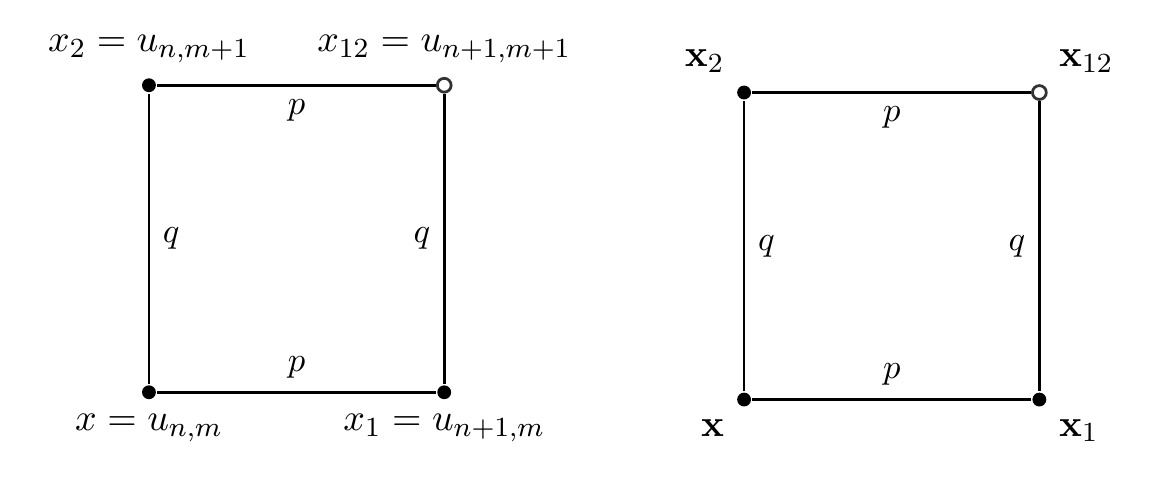}
}
 \caption{The \PDeltaE\ is defined on the simplest quadrilateral (a square).
 single equation or scalar case (left), system or vector case (right).}
 \label{fig:square}
\end{figure}

% fixed u instead of x 09/16/2019
To settle on notation, let us first consider a single {\it scalar} \PDeltaE,
\begin{equation}
 \label{eq:PDeltaE_Fu}
 {\mathcal{F}}(u_{n,m}, u_{n+1,m}, u_{n,m+1}, u_{n+1,m+1}; p, q) = 0,
\end{equation}
which is defined on a 2-dimensional quad-graph as shown in Fig.~\ref{fig:square} (left).
\vskip 1pt
% fixed indented 09/16/2019
\indent
The one field variable $x \equiv u_{n,m}$ depends on lattice variables $n$ and $m.$
A shift of $x$ in the horizontal direction (the one-direction) is denoted by $x_1 \equiv u_{n+1,m}$.
A shift in the vertical or two-direction by $x_2 \equiv u_{n,m+1}$ and a shift in both directions by
$x_{12} \equiv u_{n+1,m+1}.$
Furthermore, the nonlinear function ${\mathcal{F}}$ depends on the lattice parameters $p$ and $q$
which correspond to the edges of the quadrilateral.
In our simplified notation \eqref{eq:PDeltaE_Fu} is replaced by
\begin{equation}
 \label{eq:PDeltaE_Fx}
 {\mathcal{F}}(x, x_{1}, x_{2}, x_{12}; p, q) = 0.
\end{equation}
Alternate notations are used in the literature.
For instance, many authors denote $(x, x_1, x_2, x_{12})$ by $(x, {\tilde{x}},{\hat{x}},{\hat{\tilde{x}}})$
while others use $(x_{00}, x_{10}, x_{01}, x_{11})$.

As a well-studied example, consider the integrable lattice version of the potential Korteweg-de Vries equation
\cite{ref:NijhoffCapel1995}, written in the various notations:
% fixed throughout 09/16/2019
\begin{subequations}
 \label{eq:intropKdV}
 \begin{align}
 (p - q + u_{n,m+1} - u_{n+1,m})(p + q - u_{n+1,m+1} + u_{n,m}) &= p^2 - q^2, \\
 (p - q + u_{01} - u_{10})(p + q - u_{11} + u_{00}) &= p^2 - q^2, \\
 (p - q + {\hat{u}} - {\tilde{u}})(p + q - {\hat{\tilde{u}}} + u) &= p^2 - q^2,\\
  \intertext{or, in the notation used throughout this chapter,}
  (p - q + x_2 - x_1)(p + q - x_{12} + x) &= p^2 - q^2. \label{eq:simpleKdV}
 \end{align}
\end{subequations}
When dealing with systems of \PDeltaEs, instead of having one field variable $u_{n,m}$, there are multiple field variables,
e.g., $u_{n,m}, v_{n,m},$ and $w_{n,m},$ which we will denote by $x, y, z$.
Consequently, the scalar equation \eqref{eq:PDeltaE_Fx} is replaced by a multi-component system involving a {\it vector} function
% fixed singular variable and its instead of their 09/16/2019
${\mathbfcal{F}}$ which depends on field variable ${\bf x} \equiv (x, y, z)$ and its shifts denoted by
${\bf x}_1, {\bf x}_2,$ and ${\bf x}_{12}.$
Again, we restrict ourselves to equations defined on the quadrilateral depicted in Fig.~\ref{fig:square} (right)
which is the ``vector" version of the figure on the left.
We assume that the initial values (indicated by solid circles) for ${\bf x}, {\bf x}_1$ and ${\bf x}_2$
can be specified and that the value of ${\bf x}_{12}$ (indicated by an open circle) can be uniquely determined.
To achieve this we require that ${\mathbfcal{F}}$ is {\it affine linear} (multi-linear) in the field variables.
Eq.\ \eqref{eq:intropKdV} is an example of a class \cite{ref:Adler2003} of scalar \PDeltaEs\
% fixed chapter instead of paper 09/16/2019
which are consistent around the cube, a property that plays an important role in this chapter.

% fixed added of a system of PDeltaEs  09/16/2019
As an example of a system of \PDeltaEs, consider the Schwarzian Boussinesq system
\cite{ref:Hietarinta2011,ref:Nijhoff1997,ref:Nijhoff1999},
 \begin{subequations}
  \label{E:sBouss}
  \begin{gather}
   z y_1 - x_1 + x = 0, \;\;
   z y_2 - x_2 + x = 0,
   \;\; \text{ and }
   \label{E:sBoussE}
 \\
   (z_1 - z_2) z_{12} - \frac{z}{y} \left(p y_1 z_2 - q y_2 z_1\right)  = 0,
   \label{E:sBoussF}
  \end{gather}
 \end{subequations}
where, for simplicity of notation, $p^3$ and $q^3$ were replaced with $p$ and $q,$ respectively.
Eq.\ \eqref{E:sBoussF} is relating the four corners of the quadrilateral.
Both equations in \eqref{E:sBoussE} are not defined on the full quadrilateral.
Each is restricted to a single edge of the quadrilateral.
The first equation is defined on the edge connecting ${\bf x}$ and ${\bf x}_1$;
the second on the edge connecting ${\bf x}$ and ${\bf x}_2.$

% fixed is inherently 09/16/2019
As we will see in Section~\ref{sec:sBoussLaxPairs}, equations like \eqref{eq:intropKdV} and \eqref{E:sBouss} are very special.
Indeed, they are multi-dimensionally consistent; a property which is inherently connected to the existence and derivation
of Lax pairs.

% fixed remove partial difference equations
As in the case of nonlinear partial differential equations (PDEs), one of the fundamental characterizations of {\em integrable}
nonlinear \PDeltaEs\ is the existence of a Lax pair, i.e., an associated matrix system of two
{\em linear} difference equations for an auxiliary vector-valued function.
The original nonlinear \PDeltaE\ then arises by expressing the compatibility condition of that linear system via a commutative
diagram.

% fixed added for a discrete nonlinear Schr\"{o}dinger equation 09/16/2019
Lax pairs for \PDeltaEs\ first appeared in work by Ablowitz and Ladik
\cite{ref:AblowitzLadik1976,ref:AblowitzLadik1977} for a discrete nonlinear Schr\"{o}dinger equation,
and subsequently in \cite{ref:Nijhoff1983} for other equations.
% fixed removed hyphens, removed in [] 06/17/2019 
The existence (and construction) of a Lax pair is closely related to the so-called 
{\em consistency around the cube} (CAC) property which was (much later) proposed
independently by Nijhoff \cite{ref:Nijhoff2002} and Bobenko and Suris \cite{ref:BobenkoSuris2002}.  
CAC is a special case of {\em multi-dimensional consistency} which is nowadays used as a key criterion to
% fixed used \PDeltaEs 09/16/2019
define integrability of \PDeltaEs.

In contrast to the PDE case, there exists a straightforward, algorithmic
approach to derive Lax pairs \cite{ref:BobenkoSuris2002,ref:Nijhoff2002} for
scalar \PDeltaEs\ that are consistent around the cube, i.e., 3D consistent.
The algorithm was presented in \cite{ref:Bridgman2018,ref:Bridgmanetal2013} and has been
implemented in {\sc Mathematica} \cite{ref:Bridgman2015}.

The implementation of the algorithm for systems of \PDeltaEs\ \cite{ref:Bridgman2018,ref:Bridgmanetal2013} is more subtle,
in particular, when edge equations are present in the systems.
In the latter case, the algorithm produces gauge equivalent Lax matrices which depend on the way the edge
constraints are dealt with.
As illustrated for \eqref{E:sBouss}, incorporating the edge constraints into the calculation of Lax pairs produces
$3$ by $3$ matrices.
Not using the edge constraints also leads to valid Lax pairs involving $4$ by $4$ matrices
which are gauge-like equivalent with their $3$ by $3$ counterparts.
This was first observed \cite{ref:BridgmanHereman2018} when computing Lax pairs for systems of \PDeltaEs\
presented in Zhang {\em et al.} \cite{ref:Zhangetal2012}.
% fixed collaborators instead of collaborations, plural collaborators have  09/16/2019
Using the so-called direct linearization method, Zhang and collaborators have obtained $4$ by $4$ Lax matrices for
generalizations of Boussinesq systems derived by Hietarinta \cite{ref:Hietarinta2011}.
In Section~\ref{sec:sBoussApplic} we show the gauge-like transformations that connect these Lax matrices
with the smaller size ones presented in \cite{ref:Bridgman2018,ref:BridgmanHereman2018}.

To keep the article self-contained, in the next section we briefly discuss the concept of
{\em gauge-equivalent Lax pairs} for nonlinear PDEs and draw the analogy between the
continuous and discrete cases.
% fixed chapter instead of paper 09/16/2019
The rest of the chapter is organized as follows.
Section~\ref{sec:Boussinesqsystems} has a detailed discussion of the algorithm to compute Lax pairs with
its various options.
The leading example is the Schwarzian Boussinesq system for which various Lax pairs are computed.
The gauge and gauge-like equivalences of these Lax matrices is discussed in Section~\ref{sec:sBoussGauge}.
In Section~\ref{sec:sBoussApplic} the algorithm is applied to the generalized Hietarinta systems featured in
\cite{ref:Zhangetal2012}.
A summary of the results is given in Section~\ref{sec:sBoussSummary}.
The chapter ends with a brief discussion of the software implementation and conclusions in
Section~\ref{sec:DiscussionConclusions}.

% fixed lower cases 09/16/2019
% SECTION 2
\section{Gauge equivalence of Lax pairs for PDEs and \PDeltaEs\ }
\label{sec:gaugeequivalence}
In this section we show the analogy between Lax pairs for continuous equations (PDEs)
and lattice equations (\PDeltaEs).
We also introduce the concept of gauge equivalence in both cases.
\subsection{Lax pairs for nonlinear PDEs}
\label{sec:gaugeequivalencePDEs}
A completely integrable nonlinear PDE can be associated with a system of linear PDEs in an auxiliary function $\Phi.$
The compatibility of these linear PDEs requires that the original nonlinear PDE is satisfied.

Using the matrix formalism described in \cite{ref:AKNS1974}, we can replace a given nonlinear PDE with a linear system,
\begin{equation}
 \label{E:LPcont}
 \Phi_x = {\bf{X}} \Phi
 \; \text{ and } \;
 \Phi_t = {\bf{T}} \Phi,
\end{equation}
with vector function $\Phi(x,t)$ and unknown matrices ${\bf{X}}$ and ${\bf{T}}.$
% fixed added the equations in 09/16/2019
Requiring that the equations in \eqref{E:LPcont} are compatible,
that is requiring that $\Phi_{xt} = \Phi_{tx},$ readily \cite{ref:Hickmanetal2012}
leads to the (matrix) {\it Lax equation} (also known as the zero curvature condition) to be satisfied by the {\it Lax pair}
$({\bf{X}}, {\bf{T}})$:
\begin{equation}
\label{matLaxEq}
{\mathbf{X}}_t - {\mathbf{T}}_x + [{\mathbf{X}}, {\mathbf{T}}] \;\;\dot{=}\;\; {\mathbf{0}},
\end{equation}
where $[{\mathbf{X}}, {\mathbf{T}}] := {\mathbf{X}} {\mathbf{T}} - {\mathbf{T}} {\mathbf{X}}$
is the matrix commutator and $\dot{=}$ denotes that the equation holds for solutions of the given nonlinear PDE.
Finding the {\em Lax matrices} ${\bf{X}}$ and ${\bf{T}}$ for a nonlinear PDE (or system of PDEs) is a nontrivial task
for which to date no algorithm is available.

Consider, for example, the ubiquitous Korteweg-de Vries (KdV) equation \cite{ref:AblowitzClarkson1991},
\begin{equation}
 \label{E:kdv}
 u_t + \alpha \, u u_x + u_{xxx} = 0,
\end{equation}
where $\alpha$ is any non-zero real constant.
It is well known (see, e.g., \cite{ref:Hickmanetal2012}) that
\begin{subequations}
\label{E:kdvLAX1}
\begin{equation}
 \label{M:kdvX1}
 {\bf{X}} =
  \begin{bmatrix}
   \ 0 & 1 \ \\
    & \\
   \ \lambda - \frac{1}{6} \alpha \, u & 0 \
  \end{bmatrix}
\end{equation}
and
\begin{equation}
 \label{M:kdvT1}
 {\bf{T}} =
  \begin{bmatrix}
   \ \frac{1}{6} \alpha \, u_x & -4 \lambda - \frac{1}{3} \alpha \, u \ \\
    & \\
   \ -4\lambda^2 + \frac{1}{3} \alpha \, \lambda \, u + \frac{1}{18} \alpha^2 \, u^2 + \frac{1}{6} \alpha \, u_{xx}
   & -\,\frac{1}{6}\alpha \, u_x \
  \end{bmatrix}
\end{equation}
\end{subequations}
form a Lax pair for \eqref{E:kdv}.
In this example
$ \Phi = \begin{bmatrix} \psi & \psi_x \end{bmatrix}^{\rm T} $,
where ${\rm T}$ denotes the transpose, and $\psi(x,t)$ is the scalar eigenfunction of
the Schr\"{o}dinger equation,
\begin{equation}
\label{KdVLpsi}
\psi_{xx} - (\lambda - \tfrac{1}{6} \alpha \, u) \psi = 0,
\end{equation}
with eigenvalue $\lambda$ and potential proportional to $u(x,t).$

It has been shown \cite[p.\ 22]{ref:FaddeevTakhtajan1987} that if $({\bf{X}},{\bf{T}})$ is a Lax pair,
then so is $({\tilde{\bf X}}, {\tilde{\bf T}})$ where
\begin{equation}
\label{E:PDE-Lax-Eq}
  \atdm{X} = \contG{\bf{X}} \contG^{-1} + \contG_x \contG^{-1}
  \; \text { and } \;
  \atdm{T} = \contG{\bf{T}} \contG^{-1} + \contG_t \contG^{-1},
\end{equation}
for an arbitrary invertible matrix $\contG$ of the correct size.
The above transformation comes from changing $\Phi$ in \eqref{E:LPcont}
into $\tilde{\Phi} = \contG \Phi$ and requiring that
$ \tilde{\Phi}_x = {\tilde{\bf{X}}} \tilde{\Phi}$ and
$ \tilde{\Phi}_t = {\tilde{\bf{T}}} \tilde{\Phi}.$

In physics, transformations like \eqref{E:PDE-Lax-Eq} are called {\em gauge transformations}.
Obviously, a Lax pair for a given PDE is not unique.
In fact, there exists an infinite number of Lax pairs which are {\em gauge equivalent} through \eqref{E:PDE-Lax-Eq}.

In the case of the KdV equation, for example using the gauge matrix
\begin{equation}
 \label{M:PDE-Lax-G}
 \contG = \begin{bmatrix}
 \ -i k & 1 \ \\
 & \\
 \ -1 & 0 \ \end{bmatrix},
\end{equation}
we see that \eqref{E:kdvLAX1} is gauge equivalent to the Lax pair,
\begin{subequations}
 \label{E:kdvLAX2}
 \begin{equation}
  \label{M:kdvX2}
  {\bf{\tilde{X}}} =
   \begin{bmatrix}
    \ -i k & \frac{1}{6}\alpha \, u \ \\
     & \\
    \ -1 & i k \
   \end{bmatrix}
 \end{equation}
and
 \begin{equation}
  \label{M:kdvT2}
  {\bf{\tilde{T}}} =
   \begin{bmatrix}
    \ -4 i k^3 + \frac{1}{3} i \alpha \, k \, u - \frac{1}{6} \alpha \, u_x \;\;
      & \frac{1}{3} \alpha \left( 2 k^2\, u - \frac{1}{6} \alpha \, u^2 + i k \, u_x - \frac{1}{2} u_{xx} \right) \;\; \\
      & \\
    \ -4 k^2 + \frac{1}{3} \alpha \, u \;\;
      & 4 i k^3 - \frac{1}{3} i \alpha \, k \, u + \frac{1}{6} \alpha \, u_x \
   \end{bmatrix},
 \end{equation}
\end{subequations}
where $\lambda = - k^2.$
The latter Lax matrices are complex matrices.
However, in \eqref{M:kdvX2} the eigenvalue $k$ appears in the diagonal entries which
is advantageous if one applies the Inverse Scattering Transform (IST) to solve the initial
value problem for the KdV equation.

\subsection{Lax pairs for nonlinear \PDeltaEs\ }
\label{sec:gaugeequivalencePDeltaEs}
% fixed PDeltaE instead of PDeltaEs 09/16/2019
Analogous with the definition of Lax pairs (in matrix form) for PDEs,
a Lax pair for a nonlinear \PDeltaE\ is a pair of matrices, $(L, M),$
such that the compatibility of the linear system,
\begin{equation}
\label{LP:matrix}
\psi_1 = L \psi
\; \text{ and } \;
\psi_2 = M \psi,
\end{equation}
for an auxiliary vector function $\psi$, requires that the nonlinear \PDeltaE\ is satisfied.
The crux is to find suitable {\em Lax matrices} $L$ and $M$ so that the nonlinear \PDeltaE\
can be replaced by \eqref{LP:matrix}.
\vspace*{-4mm}
\noindent
\begin{figure*}[h]
 \centering
 \centerline{
 \includegraphics[width=50mm]{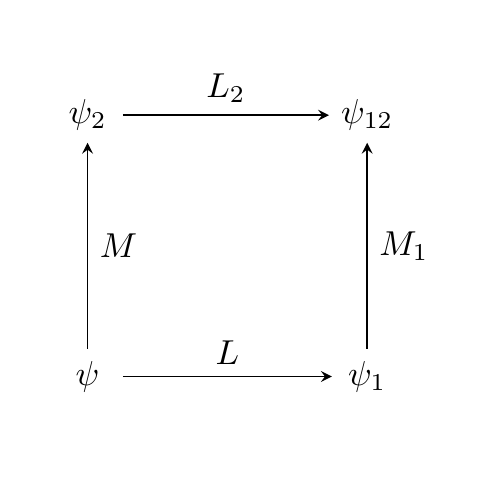}
 }
 \vspace*{-3mm}
 \caption{Commutative diagram resulting in the Lax equation.}
 \label{fig:commdiag}
\end{figure*}
\vspace*{-1mm}
\noindent
% fixed added comma 09/16/2019
As shown in the (Bianchi-type) commutative diagram depicted in Fig.~\ref{fig:commdiag},
the compatibility of \eqref{LP:matrix} can be readily expressed by shifting both sides of $\psi_1 = L \psi$
in the two-direction, i.e., $\psi_{12} = L_2 \psi_2 = L_2 M \psi$,
and shifting $ \psi_2 = M \psi$
in the one-direction, i.e., $\psi_{21} = \psi_{12} = M_1 \psi_1 = M_1 L \psi$,
and equating the results.
Hence,
\begin{equation}
\label{matLaxEqdiffeqs}
L_2 M  - M_1 L \;\; \dot{=} \;\; {0},
\end{equation}
where $\dot{=}$ denotes that the equation holds for {\em solutions} of the \PDeltaE.
% fixed left-hand instead of left hand 09/16/2019
% fixed added as this would result in  09/16/2019
In other words, the left-hand side of \eqref{matLaxEqdiffeqs} should generate the \PDeltaE\
and not be satisfied automatically as this would result in a ``fake" Lax pair.
In analogy to \eqref{matLaxEq}, equation \eqref{matLaxEqdiffeqs} is called the Lax equation
(or zero-curvature condition).

As in the continuous case, there is an infinite number of Lax matrices, all equivalent to each other under
gauge transformations \cite{ref:Bridgmanetal2013}.
Specifically, if $(L,M)$ is a Lax pair then so is $({\tilde{L}},{\tilde{M}})$ where
\begin{equation}
 \label{E:DDE-Gauge-Eq}
 {\tilde{L}} = \mc{G}_1 L \mc{G}^{-1}
 \; \text{ and } \;
 {\tilde{M}} = \mc{G}_2 M \mc{G}^{-1},
\end{equation}
for any arbitrary invertible matrix $\mc{G}$.
Gauge transformation \eqref{E:DDE-Gauge-Eq} comes from setting ${\tilde{\psi}} = \mc{G} \psi$ and
requiring that
${\tilde{\psi}}_1 = {\tilde{L}} {\tilde{\psi}}$ and ${\tilde{\psi}}_2 = {\tilde{M}} {\tilde{\psi}}.$

% fixed add not 09/16/2019
Although \eqref{E:DDE-Gauge-Eq} insures the existence of an infinite number of Lax matrices, it does not say
how to find $\mc{G}$ of any two Lax matrices (which might have been derived with different methods).
As we shall see, there are systems of \PDeltaEs\ with Lax matrices whose gauge equivalence is presently unclear.

% fixed lower cases 09/16/2019
% SECTION 3 -- ALGORITHM
\section{Derivation of Lax pairs for Boussinesq systems}
 \label{sec:Boussinesqsystems}
\subsection{Derivation of Lax Pairs for the Schwarzian Boussinesq System}
 \label{sec:sBoussLaxPairs}
\begin{figure*}[h]
  \centering
\centerline{
 \includegraphics[width=60mm]{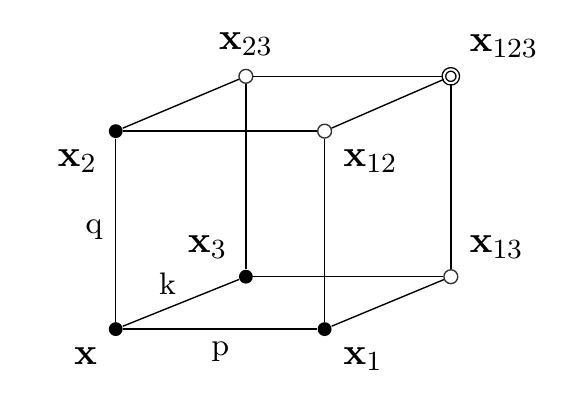}
 }
  \vspace*{-2mm}
  \caption{The system of \PDeltaEs\ holds on each face of the cube.}
  \label{fig:cube}
 \end{figure*}

\vskip 1pt
\noindent
% fixed lower cases 09/16/2019
% {\bf Consistency Around the Cube}
{\bf Consistency around the cube}
\vskip 5pt
\noindent
The key idea of multi-dimensional consistency is to
 (i) extend the planar quadrilateral (square) to a cube by artificially
  introducing a third direction (with lattice parameter $k$) as shown in
  Fig.~\ref{fig:cube},
 (ii) impose copies of the same system, albeit with different lattice parameters,
  on the different faces and edges of the cube, and
 (iii) view the cube as a three-dimensional commutative diagram for ${\bf x}_{123}.$
 % fixed cut all and parameter instead of parameters 09/16/2019
  Although not explicitly shown in Fig.~\ref{fig:cube}, parallel edges carry the
  same lattice parameter.
 \vspace*{-4mm}
  \noindent
 \vskip 0.001pt
 \indent
As shown in Fig.~\ref{fig:cube}, the planar quadrilateral is extended into the third dimension where $k$ is
the lattice parameter along the edge connecting ${\bf x}$ and ${\bf x}_3.$
Although not explicitly shown in Fig.~\ref{fig:cube}, all parallel edges carry the same lattice parameters.

With regard to \eqref{E:sBouss}, we impose that the same equations hold on all faces of the cube.
% fixed em instead of rm 09/16/2019
The equations on the {\em{bottom}} face follow from a rotation of the {\em{front}} face along the
horizontal axis connecting ${\bf x}$ and ${\bf x}_1.$
% fixed applying instead of apply and yields instead of to get, moved to eq. 09/16/2019
Therefore, applying the substitutions
${\bf x}_2 \rightarrow {\bf x}_3, \; {\bf x}_{12} \rightarrow {\bf x}_{13}, \text{ and } q \rightarrow k$
to \eqref{E:sBouss}, yields
\begin{subequations}
  \label{E:sBoussbottom}
  \begin{gather}
   z y_1 - x_1 + x = 0, \;\;
   z y_3 - x_3 + x = 0,
   \;\; \text{ and }
   \label{E:sBoussEbottom}
 \\
   (z_1 - z_3) z_{13} - \frac{z}{y} \left(p y_1 z_3 - k y_3 z_1 \right)  = 0,
   \label{E:sBoussFbottom}
  \end{gather}
 \end{subequations}
which visually corresponds to ``folding" the front face down into the bottom face.

% fixed em instead of rm on left and rm on front 09/16/2019
Likewise, the equations on the {\em{left}} face can be obtained via a rotation of the {\rm{front}} face along the
vertical axis connecting ${\bf x}$ and ${\bf x}_2$ (over ninety degrees counterclockwise from a bird's eye view).
% fixed moved to eq. before yielding 09/16/2019
This amounts to applying the substitutions
${\bf x}_1 \rightarrow {\bf x}_3, \; {\bf x}_{12} \rightarrow {\bf x}_{23}, \text{ and }  p \rightarrow k$
to \eqref{E:sBouss}, yielding
\begin{subequations}
  \label{E:sBoussleft}
  \begin{gather}
   z y_3 - x_3 + x = 0, \;\;
   z y_2 - x_2 + x = 0,
   \;\; \text{ and }
   \label{E:sBoussEleft}
 \\
   (z_3 - z_2) z_{23} - \frac{z}{y} \left(k y_3 z_2 - q y_2 z_3 \right)  = 0.
   \label{E:sBoussFleft}
  \end{gather}
 \end{subequations}
The equations on the {\em{back}} face follow from a shift of \eqref{E:sBouss} in the third direction letting
${\bf x} \rightarrow {\bf x}_3, \; {\bf x}_1 \rightarrow {\bf x}_{13}, \;  {\bf x}_2 \rightarrow {\bf x}_{23},
\text{ and } {\bf x}_{12} \rightarrow {\bf x}_{123}.$
% fixed em on top and right 09/16/2019
Likewise, the equations on the {\em{top}} and {\em{right}} faces follow from \eqref{E:sBoussbottom} and \eqref{E:sBoussleft}
by shifts in the two- and one-directions, respectively.

Gathering the equations (from all six faces) yields a system of 15 equations (after removing the three duplicates).
For example, the five equations that reside on the {\it bottom} face
(with corners ${\bf x}, {\bf x}_1, {\bf x}_3$, and ${\bf x}_{13})$ are
\begin{subequations}
\label{E:sBoussfloorall}
\begin{align}
   & z y_1 - x_1 + x = 0, \label{E:sBoussflooralla} \\
   & z y_3 - x_3 + x = 0, \label{E:sBoussfloorallb} \\
   & z_3 y_{13} - x_{13} + x_3 = 0, \label{E:sBoussfloorallc} \\
   & z_1 y_{13} - x_{13} + x_1 = 0, \label{E:sBoussflooralld} \\
   &(z_1 - z_3) z_{13} - \frac{z}{y}\left(p y_1 z_3 - k y_3 z_1 \right)  = 0,
   \label{E:sBoussflooralle}
  \end{align}
 \end{subequations}
 yielding the components of ${\bf x}_{13}$, namely,
 \begin{subequations}
  \label{E:sBouss13}
  \begin{align}
   x_{13} &= \frac{x_3z_1 - x_1z_3}{z_1 - z_3}, \;\;
   y_{13} = \frac{x_3 - x_1}{z_1 - z_3},
   \; \text{ and } \\
   z_{13} &= \frac{z}{y}\left(\frac{p y_1z_3 - k y_3z_1}{z_1 - z_3}\right).
  \end{align}
 \end{subequations}
Likewise, solving the equations on the front face yields the components of ${\bf x}_{12}$ and
the equations on the left face yield the components of ${\bf x}_{23}.$
The components of ${\bf x}_{123}$ can be computed using either the equations on the top face or
those on the right or back faces.

Multi-dimensional consistency around the cube of the \PDeltaE\ system requires that one can uniquely determine
${\bf x}_{123} = (x_{123}, y_{123}, z_{123})$ and that all expressions coincide, no matter which face is used
(or, equivalently, no matter which path along the cube is taken to get to the corner ${\bf x}_{123}$).
Using straightforward, yet tedious algebra, one can show \cite{ref:Bridgmanetal2013} that \eqref{E:sBouss} is
multi-dimensionally consistent around the cube.
As discussed in \cite{ref:BobenkoSuris2008}, three-dimensional consistency of a system of \PDeltaEs\ establishes
its complete integrability for it allows one to algorithmically compute a Lax pair.
\vskip 5pt
\noindent
% fixed lower cases 09/16/2019
% {\bf Computation of Lax Pairs}
{\bf Computation of Lax pairs}
\vskip 5pt
\noindent
The derivation of a Lax pair for \eqref{E:sBouss} starts with introducing {\em projective} variables
$\fnum{x}, \fnum{y}, \fnum{z}, \fden{x}, \fden{y},$ and $\fden{z}$ by
 \begin{equation}
  \label{E:xyzSubs}
   x_3 = \frac{\fnum{x}}{\fden{x}}, \;\;
   y_3 = \frac{\fnum{y}}{\fden{y}},
   \; \text{ and } \;
   z_3 = \frac{\fnum{z}}{\fden{z}}.
 \end{equation}
Note that the numerators and denominators of \eqref{E:sBouss13} are linear in $x_3, y_3,$ and $z_3.$
The above fractional transformation allows one to make the top and bottom of $x_{13}, y_{13},$ and $z_{13}$
linear in the projective variables (in the same vein as using fractional transformation to linearize Riccati
equations).

Substitution of \eqref{E:xyzSubs} into \eqref{E:sBouss13} yields
\begin{subequations}
  \label{E:sBouss13projvars}
  \begin{align}
   x_{13} &= \frac{z_1 H f - x_1 F h }{F (z_1 H - h)}, \;\;
   y_{13} = -\, \frac{H (x_1 F - f)}{F (z_1 H - h)}, \; \text{ and } \\
   z_{13} &= -\, \frac{z}{y}\left(\frac{k z_1 H g - p y_1 G h}{G (z_1 H - h)}\right).
  \end{align}
 \end{subequations}
Achieving the desired linearity requires $F = G = H.$
Then, \eqref{E:sBouss13projvars} becomes
\begin{subequations}
  \label{E:sBouss13projvarssimp}
  \begin{align}
   x_{13} &= \frac{z_1\fnum{x} - x_1\fnum{z}}{z_1\fden{x} - \fnum{z}}, \;\;
   y_{13} = -\, \frac{x_1\fden{x} - \fnum{x}}{z_1\fden{x} - \fnum{z}}, \; \text{ and } \\
   z_{13} &= -\, \frac{z}{y}\left(\frac{k z_1\fnum{y} - p y_1\fnum{z}}{z_1 \fden{x} - \fnum{z}}\right),
  \end{align}
 \end{subequations}
with
\begin{equation}
  \label{E:xyzSubs3}
   x_3 = \frac{\fnum{x}}{\fden{x}}, \;\;
   y_3 = \frac{\fnum{y}}{\fden{x}},
   \; \text{ and } \;
   z_3 = \frac{\fnum{z}}{\fden{x}}.
 \end{equation}
How one deals with the remaining variables $f, g, h,$ and $F,$ leads to various
alternatives for Lax matrices.

% fixed lower cases 09/16/2019
\subsubsection{The first alternative}
\label{sec:firstAlt1}
% CHOICE 1
{\bf Choice 1.}$\;$ Note that the edge equation \eqref{E:sBoussfloorallb}
imposes an additional constraint on \eqref{E:xyzSubs3}.
Indeed, solving \eqref{E:sBoussfloorallb} for $x_3$ in terms of $y_3$ yields
% fixed period instead of comma 09/16/2019
\begin{equation}
\label{E:x3y3rel1}
x_3 = z y_3 + x.
\end{equation}
% fixed added x_3 piece 09/16/2019
Using $x_3 = \frac{\fnum{x}}{\fden{x}}$ and $y_3 = \frac{\fnum{y}}{\fden{x}}$
one can eliminate $\fnum{x}$ since $\fnum{x} = x \fden{x} + z \fnum{y}$.
Eqs.\ \eqref{E:xyzSubs3} and \eqref{E:sBouss13projvarssimp} then become
\begin{equation}
  \label{E:xyzSubsx3}
   x_3 = \frac{x \fden{x} + z \fnum{y}}{\fden{x}}, \;\;
   y_3 = \frac{\fnum{y}}{\fden{x}},
   \; \text{ and } \;
   z_3 = \frac{\fnum{z}}{\fden{x}},
 \end{equation}
and
\begin{subequations}
 \label{E:sBouss13projA}
  \begin{align}
   x_{13} & = \frac{\fnum{x}_1}{\fden{x}_1}
    = \frac{x_1 \fden{x}_1 + z_1\fnum{y}_1}{\fden{x}_1}
    = \frac{x z_1\fden{x} + z z_1\fnum{y} - x_1\fnum{z}}{z_1\fden{x} - \fnum{z}},
    \label{E:sBoussPx13} \\
   y_{13} & = \frac{\fnum{y}_1}{\fden{x}_1}
    = \frac{(x - x_1)\fden{x} + z\fnum{y}}{z_1\fden{x} - \fnum{z}},
    \label{E:sBoussPy13} \\
   z_{13}  & = \frac{\fnum{z}_1}{\fden{x}_1}
   = -\, \frac{z (k z_1\fnum{y} - p y_1\fnum{z})}{y (z_1\fden{x} - \fnum{z})},
   \label{E:sBoussPz13}
  \end{align}
 \end{subequations}
where $\fden{x}, \fnum{y},$ and $\fnum{z}$ are independent (and remain undetermined).

Then, \eqref{E:sBoussPy13} and \eqref{E:sBoussPz13} can be split by setting
\begin{subequations}
 \label{E:sBoussf1a}
 \begin{align}
  \fden{x}_1 =\ & t \, ( z_1\fden{x} - \fnum{z} ), \\
  \fnum{y}_1 =\ & t \, \big( (x - x_1)\fden{x} + z\fnum{y} \big), \\
  \fnum{z}_1 =\ & t \left(- \frac{z}{y} \,(k z_1\fnum{y} - p y_1\fnum{z})\right),
 \end{align}
\end{subequations}
where $t({\bf x}, {\bf x}_1; p, k)$ is a scalar function still to be determined.
One can readily verify that \eqref{E:sBoussPx13} is identically satisfied.

If we define
$ \psi_{\rm a} := \begin{bmatrix}
  \fden{x} & \fnum{y} & \fnum{z} \end{bmatrix}^{\rm T}$,
  then
$\left(\psi_{\rm a} \right)_1 = \begin{bmatrix}
  \fden{x}_1 & \fnum{y}_1 & \fnum{z}_1 \end{bmatrix}^{\rm T}$,
we can write \eqref{E:sBoussf1a} in matrix form
$ \left( \psi_{\rm a} \right)_1 = L_{\rm a} \psi_{\rm a},$ with
\begin{equation}
 \label{M:sBoussLa}
 L_{\rm a} = t \, L_{{\rm a}, {\rm core}}
   := t \begin{bmatrix}
        z_1 & 0 & -1 \\
        & & \\
        x - x_1 & z & 0 \\
        & & \\
        0 & -\,\frac{k zz_1}{y} & \frac{p zy_1}{y}
     \end{bmatrix}.
\end{equation}
The partner matrix $M_{\rm a}$ of the Lax pair,
\begin{equation}
 \label{M:sBoussMa}
  M_{\rm a} = s \, M_{{\rm a}, {\rm core}}
    := s \begin{bmatrix}
        z_2 & 0 & -1 \\
        & & \\
        x - x_2 & z & 0 \\
        & & \\
        0 & -\,\frac{k zz_2}{y} & \frac{q zy_2}{y}
     \end{bmatrix},
\end{equation}
comes from substituting \eqref{E:xyzSubsx3} into the five equations (similar to \eqref{E:sBoussfloorall})
for the left face of the cube.
Formally, this amounts to replacing all indices $1$ by $2$ and $p$ by $q$ in $L_{{\rm a}, {\rm core}}$
(see, e.g., \cite{ref:Bridgmanetal2013} for details).
In subsequent examples, the partner matrices $(M)$ will no longer be shown.

Using the same terminology as in \cite{ref:Bridgmanetal2013},
$L_{{\rm a}, {\rm core}}$ and $M_{{\rm a}, {\rm core}}$ are the ``core''
of the Lax matrices $L_{\rm a}$ and $M_{\rm a},$ respectively.
The label ``a" on $\psi_{\rm a}, L_{\rm a},$ and $M_{\rm a}$ is added to
differentiate the entries within each family of Lax matrices (up to trivial
permutations of the components).
In what follows, alternative choices will be labeled with ``b," ``A", ``B," etc.
These matrices come from alternate ways of treating the edge equations.

The functions $t({\bf x}, {\bf x}_1; p, k)$ and $s({\bf x}, {\bf x}_2; q, k)$
can be computed algorithmically as shown in \cite{ref:Bridgmanetal2013} or by
using the Lax equation \eqref{matLaxEqdiffeqs} directly, as follows,
\begin{equation}
\label{E:laxEQLM1}
\left( t L_{\rm core} \right)_2 (s M_{\rm core}) - \left( s M_{\rm core} \right)_1 (t L_{\rm core})
\!=\! (s\, t_2) \left( L_{\rm core} \right)_2 M_{\rm core} - (t \, s_1) \left( M_{\rm core} \right)_1 L_{\rm core}
\,\dot{=}\, 0,
\end{equation}
which implies that
\begin{equation}
\label{E:laxEQLM2}
\frac{s \, t_2}{t \, s_1} \left( L_{\rm core} \right)_2 M_{\rm core}
\;\;\dot{=}\;\; \left( M_{\rm core} \right)_1 L_{\rm core}.
\end{equation}
After replacing $L_{\rm core}$ and $M_{\rm core}$ by $L_{{\rm a},{\rm core}}$
and $M_{{\rm a},{\rm core}}$ from \eqref{M:sBoussLa} and \eqref{M:sBoussMa},
respectively, in \eqref{E:laxEQLM2}, one gets
\begin{equation}
 \label{E:sBoussst}
 \frac{s \, t_2}{t \, s_1} \;\; \dot{=}\;\; \frac{z_1}{z_2},
\end{equation}
which has an infinite family of solutions.
% fixed left-hand instead of left hand 09/16/2019
Indeed, the left-hand side of \eqref{E:sBoussst} is invariant under the change
\begin{equation}
 \label{E:stChange}
t \rightarrow \frac{i_1}{i} \, t, \quad s \rightarrow \frac{i_2}{i} \, s,
\end{equation}
where $i({\bf x})$ is an arbitrary function and $i_1$ and $i_2$ denote the shifts of $i$
in the one- and two-direction, respectively.
One can readily verify that \eqref{E:sBoussst} is satisfied by, for example,
$ s = t = \frac{1}{z}. $
Then, for $\psi_{\rm a} = \begin{bmatrix} \fden{x} & \fnum{y} & \fnum{z} \end{bmatrix}^{\rm T}$,
the Lax matrix $L_{\rm a}$ in \eqref{M:sBoussLa} becomes
\begin{equation}
 \label{M:sBoussLaxPairA}
  L_{\rm a} = \frac{1}{z} \begin{bmatrix}
        z_1 & 0 & -1 \\
        & & \\
        x - x_1 & z & 0 \\
        & & \\
        0 & -\,\frac{k zz_1}{y} & \frac{p zy_1}{y}
     \end{bmatrix}.
\end{equation}
% CHOICE 2
{\bf Choice 2.}$\;$ Solving the edge equation \eqref{E:sBoussfloorallb} for $y_3$,
\begin{equation}
  \label{E:x3y3rel2}
  y_3 = \frac{x_3 - x}{z},
 \end{equation}
and using \eqref{E:xyzSubs3} yields $\fnum{y} = -\, \frac{x\fden{x} - \fnum{x}}{z}.$
Then \eqref{E:xyzSubs3} becomes
\begin{equation}
\label{E:xyzSubsOpt2}
x_3 = \frac{\fnum{x}}{\fden{x}}, \;\;
y_3 = -\, \frac{x \fden{x} - \fnum{x} }{z\fden{x}},
\; \text{ and } \;
z_3 = \frac{\fnum{z}}{\fden{x}},
\end{equation}
where $\fden{x}, \fnum{x},$ and $\fnum{z}$ are now independent.
Eqs.\ \eqref{E:sBouss13} now become
\begin{subequations}
 \label{E:sBouss13projB}
  \begin{align}
   x_{13} &= \frac{\fnum{x}_1}{\fden{x}_1}
   = \frac{z_1\fnum{x} - x_1\fnum{z}}{z_1\fden{x} - \fnum{z}}, \label{E:sBouss13Bx} \\
   y_{13} &= \frac{\fnum{y}_1}{\fden{x}_1}
   = -\, \frac{x_1\fden{x}_1 - \fnum{x}_1 }{z_1 \fden{x}_1}
   = -\, \frac{x_1\fden{x} - \fnum{x} }{z_1\fden{x} - \fnum{z}},
   \label{E:sBouss13By} \\
   z_{13} &= \frac{\fnum{z}_1}{\fden{x}_1}
   = \frac{k xz_1\fden{x} - k z_1\fnum{x} + p zy_1\fnum{z}}{y(z_1\fden{x} - \fnum{z})}, \label{E:sBouss13Bz}
  \end{align}
 \end{subequations}
which can be split by selecting
\begin{subequations}
 \label{E:sBoussf1b}
 \begin{align}
  \fden{x}_1 =\ & t \, ( z_1\fden{x} - \fnum{z} ), \\
  \fnum{x}_1 =\ & t \, ( z_1\fnum{x} - x_1\fnum{z} ), \\
  \fnum{z}_1 =\ & t \left( \frac{1}{y} \, (k xz_1\fden{x} - k z_1\fnum{x} + p zy_1\fnum{z}) \right),
 \end{align}
\end{subequations}
where $t({\bf x}, {\bf x}_1; p, k)$ is a scalar function still to be determined.
Note that \eqref{E:sBouss13By} is identically satisfied.

If we define
$ \psi_{\rm b} := \begin{bmatrix}
  \fden{x} & \fnum{x} & \fnum{z} \end{bmatrix}^{\rm T}$,
  then
$\left(\psi_{\rm b} \right)_1 = \begin{bmatrix}
  \fden{x}_1 & \fnum{x}_1 & \fnum{z}_1 \end{bmatrix}^{\rm T}$,
we can write \eqref{E:sBoussf1b} in matrix form,
$\left( \psi_{\rm b} \right)_1 = L_{\rm b} \psi_{\rm b},$
with
\begin{equation}
 \label{M:sBoussLb}
 L_{\rm b} = t \, L_{{\rm b}, {\rm core}}
   := \frac{1}{z} \begin{bmatrix}
        z_1 & 0 & -1 \\
        & & \\
        0 & z_1 & -x_1 \\
        & & \\
        \frac{k xz_1}{y} & -\,\frac{k z_1}{y} & \frac{p zy_1}{y}
     \end{bmatrix},
\end{equation}
where we substituted $t = \frac{1}{z}$ which was computed the same way as in Choice 1.

Thus, within this first alternative, the two choices of representing the edge constraint result in
minimally-sized Lax matrices which we call \textit{representative} Lax matrices for the \PDeltaE.
%% Second Alternative subsection
%% fixed lower cases 09/16/2019
\subsubsection{The second alternative}
\label{sec:firstAlt2}
As a second alternative in the algorithm, we do not use the edge equations to replace
\eqref{E:xyzSubs3} by \eqref{E:xyzSubsx3} or \eqref{E:xyzSubsOpt2}.
Instead, we incorporate the edge equations \eqref{E:sBoussEbottom} into {\it all three} equations of
\eqref{E:sBouss13}.
% fixed added comma 09/16/2019
Using \eqref{E:sBoussEbottom}, we replace $x_3$ and $x_1$ in $x_{13}$ and $y_{13}$, and $y_1$ and $y_3$ in $z_{13},$
yielding
 \begin{subequations}
  \label{E:xyz13m}
  \begin{align}
   \tilde{x}_{13} &= x + \frac{z(y_3z_1 - y_1z_3)}{z_1 - z_3}
          = \frac{x(z_1 - z_3) + z(y_3z_1 - y_1z_3)}{z_1 - z_3},
   \\
   \tilde{y}_{13} &= \frac{z(y_3 - y_1)}{z_1 - z_3}, \;\; \text{ and } \;\;
   \tilde{z}_{13} = \frac{k z_1 (x - x_3) - p z_3 (x - x_1)}{y(z_1 - z_3)}.
  \end{align}
 \end{subequations}
 % fixed cut extra the 09/16/2019
 Note that each of the above expressions has the same denominator $z_1 - z_3$ as in \eqref{E:sBouss13}.
 Thus, in principle, any equation from \eqref{E:sBouss13} could be replaced by the matching equation from \eqref{E:xyz13m}.
\vskip 10pt
\noindent
{\bf Choice 1.}$\;$ If we take $x_{13}$, $\tilde{y}_{13}$ and $z_{13}$, then substitution of \eqref{E:xyzSubs3} yields
 \begin{subequations}
 \label{E:sBouss13projC}
  \begin{align}
   x_{13} &= \frac{\fnum{x}_1}{\fden{x}_1}
     = \frac{z_1\fnum{x} - x_1\fnum{z}}{z_1\fden{x} - \fnum{z}}, \;\;
   \tilde{y}_{13} = \frac{\fnum{y}_1}{\fden{x}_1}
     = -\, \frac{z(y_1\fden{x} - \fnum{y})}{z_1\fden{x} - \fnum{z}},
   \;\; \text{ and }
   \\
   z_{13} &= \frac{\fnum{z}_1}{\fden{x}_1}
     = -\, \frac{z (k z_1\fnum{y} - p y_1 \fnum{z})}{y (z_1\fden{x} - \fnum{z})}.
  \end{align}
 \end{subequations}
Hence, we set
\begin{subequations}
 \label{E:sBoussf1c}
 \begin{align}
  \fden{x}_1 =\ & t \, ( z_1\fden{x} - \fnum{z} ), \\
  \fnum{x}_1 =\ & t \, ( z_1\fnum{x} - x_1\fnum{z} ), \\
  \fnum{y}_1 =\ & t \, \big( -(zy_1\fden{x} - z\fnum{y}) \big),
  \\
  \fnum{z}_1 =\ & t \left( -\frac{z}{y}\,\big(k z_1 \fnum{y} - p y_1\fnum{z}\big) \right),
 \end{align}
\end{subequations}
 where $t({\bf x}, {\bf x}_1; p, k)$ is a scalar function still to be determined.

 Defining
 $\psi := \begin{bmatrix}
  \fden{x} & \fnum{x} & \fnum{y} & \fnum{z}
  \end{bmatrix}^{\rm T}$,
 yields
 $\psi_1 = \begin{bmatrix}
  \fden{x}_1 & \fnum{x}_1 & \fnum{y}_1 & \fnum{z}_1 \end{bmatrix}^{\rm T}$.
 So, we can write \eqref{E:sBoussf1c} as $\psi_1 = L_A \psi,$ with
 \begin{equation}
 \label{M:sBoussLc}
  L_{\rm A} = t \, L_{\rm core}
   := \frac{1}{z} \, \begin{bmatrix}
        z_1 & 0 & 0 & -1 \\
        & & \\
        0 & z_1 & 0 & -x_1 \\
        & & \\
        -zy_1 & 0 & z & 0 \\
        & & \\
        0 & 0 & -\,\frac{k zz_1}{y} & \frac{p zy_1}{y}
     \end{bmatrix},
 \end{equation}
where we have used \eqref{E:laxEQLM1} to get $t = \frac{1}{z}$.

Though not obvious at first glance, $L_{\rm a}$ in \eqref{M:sBoussLa} follows from $L_{\rm A}$ after removing the second row
and second column and replacing $z y_1$ by $x_1 - x$ using \eqref{E:sBoussEbottom}.
Matrices \eqref{M:sBoussLc} and $M_{\rm A}$ (obtained from $L_{\rm A}$ by replacing indices
% fixed larger size instead of higher dimension, larger-sized instead of larger size 09/16/2019
$1$ by $2$ and $p$ by $q$ in $L_{\rm A}$) are a valid Lax pair despite being of larger size than $(L_{\rm a}, M_{\rm a}).$
We refer to these larger-sized Lax matrices as \textit{extended} Lax matrices of \PDeltaEs.
\vskip 10pt
\noindent
% fixed added substitution of and replaced yield by yields 09/16/2019
{\bf Choice 2.}$\;$ If we work with $x_{13}$, $\tilde{y}_{13}$ and $\tilde{z}_{13}$
then substitution of \eqref{E:xyzSubs3} yields
 \begin{subequations}
  \label{E:sBouss13projD}
  \begin{align}
   x_{13} &= \frac{\fnum{x}_1}{\fden{x}_1} =
    \frac{z_1\fnum{x} - x_1\fnum{z}}{z_1\fden{x} - \fnum{z}}, \;\;
   \tilde{y}_{13} = \frac{\fnum{y}_1}{\fden{x}_1} =
    -\, \frac{zy_1\fden{x} - z\fnum{y}}{z_1\fden{x} - \fnum{z}},
    \;\; \text{ and }
   \\
   \tilde{z}_{13} &= \frac{\fnum{z}_1}{\fden{x}_1} =
    \frac{k xz_1\fden{x} - k z_1\fnum{x} - p (x - x_1)\fnum{z}}{y(z_1\fden{x} - \fnum{z})}.
  \end{align}
 \end{subequations}
Setting
\begin{subequations}
 \label{E:sBoussf1d}
 \begin{align}
  \fden{x}_1 =\ & t \, ( z_1\fden{x} - \fnum{z} ), \\
  \fnum{x}_1 =\ & t \, ( z_1\fnum{x} - x_1\fnum{z} ), \\
  \fnum{y}_1 =\ & t \, \big( -(zy_1\fden{x} - z\fnum{y}) \big),
  \\
  \fnum{z}_1 =\ & t \left( \frac{1}{y} \,\big( k xz_1\fden{x} - k z_1\fnum{x} - p (x - x_1) \fnum{z} \big) \right),
 \end{align}
\end{subequations}
and defining $\psi$ as in Choice 1, yields
\begin{equation}
 \label{M:sBoussLd}
 L_{\rm B} =  t \, L_{\rm core}
   := \frac{1}{z} \, \begin{bmatrix}
        z_1 & 0 & 0 & -1 \\
        & & \\
        0 & z_1 & 0 & -x_1 \\
        & & \\
        -zy_1 & 0 & z & 0 \\
        & & \\
        \frac{k xz_1}{y}  & -\,\frac{k z_1}{y}  & 0 & -\,\frac{p (x - x_1)}{y}
     \end{bmatrix},
 \end{equation}
since $t = \frac{1}{z}$.
Thus, $L_{\rm B}$ is another extended Lax pair for the Schwarzian Boussinesq system.
Note that $L_{\rm B}$ reduces to $L_{\rm b}$ in \eqref{M:sBoussLb}
by removing the third row and third column and
replacing $x_1 - x$ by $z y_1$ based on \eqref{E:sBoussEbottom}.
\vskip 10pt
\noindent
% fixed other instead of others 09/16/2019
{\bf Choices 3 and 4.}$\;$ Repeating the process with other combinations of solutions
from \eqref{E:sBouss13} and \eqref{E:xyz13m} results in the following Lax matrices,
 \begin{equation}
  \label{M:sBoussXY}
  L_{\rm C} = \frac{1}{z} \, \begin{bmatrix}
     z_1 & 0 & 0 & -1\\  xz_1 & 0 & zz_1 & -(x + zy_1)\\
     -zy_1 & 0 & z & 0\\ 0 & 0 & -\,\frac{k zz_1}{y}  & \frac{p zy_1}{y}
  \end{bmatrix},
 \end{equation}
 and
 \begin{equation}
  \label{M:sBoussZ}
  L_{\rm D} = \frac{1}{z} \, \begin{bmatrix}
     z_1 & 0 & 0 & -1\\  0 & z_1 & 0 & -x_1\,\\
     -x_1 & 1 & 0 & 0\\  \frac{k xz_1}{y} & -\,\frac{k z_1}{y} & 0 & \frac{p (x_1 - x)}{y}
  \end{bmatrix},\\
 \end{equation}
from working with $\tilde{x}_{13}, \tilde{y}_{13}, z_{13}$ and $x_{13}, y_{13}, \tilde{z}_{13}$, respectively.

Obviously $L_{\rm C}$ and $L_{\rm D}$ are trivially related to $L_{\rm a}$ and $L_{\rm b}$, respectively.
Indeed, remove the second row and second column in $L_{\rm C}$ and the third row and third
column in $L_{\rm D}$ and use $z y_1 = x_1 - x$ to get $L_{\rm a}$ and $L_{\rm b}$, respectively.

All other combinations of pieces of \eqref{E:sBouss13} and \eqref{E:xyz13m} do {\it not} lead to matrices
that satisfy the defining equation \eqref{matLaxEqdiffeqs}.

The second alternative leads to extended Lax matrices but they are not always trivial extensions of the
representative Lax matrices.
In other words, smaller-size matrices do not necessarily follow from the larger-size matrices by simply
removing rows and columns.
Furthermore, if the representative Lax matrices were not known it would not be obvious which rows and columns
should be removed.
As shown in Section \ref{sec:sBoussSummary}, some edge-constrained systems have extended Lax matrices
whose ``equivalence" to a representative Lax matrix is non-trivial.

In the next section we carefully investigate the connections between the Lax matrices computed with the two
alternatives and various choices above.

% fixed lower cases 09/16/2019
% SECTION 4 -- GAUGE AND GAUGE LIKE EQUIVALENCES
\section{Gauge and gauge-like equivalences of Lax pairs}
\label{sec:sBoussGauge}
% fixed put one in italics 09/16/2019
As discussed in Section \ref{sec:gaugeequivalence}, if there exists {\it one} pair of Lax matrices
% fixed added of such pairs 09/16/2019
for a given system of \PDeltaEs\ then there is an {\it infinite} number of such pairs,
all equivalent to each other under discrete gauge transformations of type
\eqref{E:DDE-Gauge-Eq} involving a square matrix $\mc{G}.$

Given two distinct Lax pairs with matrices of the same size (no matter how they were computed),
it has yet to be shown if there exists a gauge transformation relating them.
For edge-constrained systems, the derivation of the gauge transformation between {\it representative}
Lax matrices is straightforward.
However, the derivation of a gauge-like transformation between Lax matrices of different sizes
(resulting from the application of different methods) is nontrivial.
In this section we look at gauge and gauge-like transformations in more detail.

% fixed lower cases 09/16/2019
%%  sub-section Gauge Equivalence
\subsection{Gauge equivalence}
\label{sec:GaugeEquiv}
In Section~\ref{sec:firstAlt1} we obtained \eqref{M:sBoussLaxPairA} and \eqref{M:sBoussLb}, resulting from
Choices 1 and 2 of dealing with the edge constraints.
\vskip 8pt
\noindent
{\bf Example 1.}$\;$ With regard to \eqref{E:DDE-Gauge-Eq}, computing the gauge matrix $\mc{G}$ such that
\begin{equation}
\label{E:altGauge1}
L_{\rm b} = \mc{G}_1 L_{\rm a} \mc{G}^{-1}
\end{equation}
% fixed do not refer to eq instead us the gauge relationship 09/16/2019
is straightforward if we consider the implications of the gauge relationship.
% \eqref{E:DDE-Gauge-Eq}.
Indeed, multiplying \eqref{E:altGauge1} by $\psi_{\rm b},$ and using \eqref{LP:matrix} yields
 \begin{subequations}
  \begin{equation}
   \label{E:altGauge2}
   \left(\psi_{\rm b} \right)_1
    = L_{\rm b} \psi_{\rm b} = (\mc{G}_1 L_{\rm a} \mc{G}^{-1}) \psi_{\rm b}.\\
  \end{equation}
 \end{subequations}
 Hence, if we set $\dsp \psi_{\rm a} = \mc{G}^{-1} \psi_{\rm b},$ we obtain
 $ \left( \psi_{\rm b} \right)_1 = \mc{G}_1 L_{\rm a} \psi_{\rm a}
     = \mc{G}_1 \left( \psi_{\rm a} \right)_1
     = \left( \mc{G} \psi_{\rm a} \right)_1.$
 Thus, $\dsp \psi_{\rm b} = \mc{G} \psi_{\rm a}$ determines $\mc{G}$.
 Not surprisingly, the gauge matrix $\mc{G}$ depends on how we selected the components of $\psi$
 which, in turn, depends on how the edge equation \eqref{E:sBoussfloorallb} was treated.

Recall that
$ \psi_{\rm a} := \begin{bmatrix} \fden{x} & \fnum{y} & \fnum{z} \end{bmatrix}^{\rm T}$
and
$ \psi_{\rm b} := \begin{bmatrix} \fden{x} & \fnum{x} & \fnum{z} \end{bmatrix}^{\rm T}.$
Using \eqref{E:xyzSubsx3}, we get
 \begin{equation}
  \label{M:sB-GaugeAB}
  \mc{G} = \begin{bmatrix}
            1  & 0  & 0 \\
            x  & z  & 0 \\
            0  & 0  & 1
           \end{bmatrix} \;\; \text{ and } \;\;
  \mc{G}^{-1} = \begin{bmatrix}
                 1 & 0 & 0 \\
                 -\,\frac{x}{z} & \frac{1}{z} & 0 \\
                 0 & 0 & 1
                \end{bmatrix}.
 \end{equation}
Indeed,
\begin{equation}
 \label{M:sB-GaugeEquiv}
 \mc{G} \psi_{\rm a} = \begin{bmatrix}
                          1 & 0 & 0 \\
                          x & z & 0 \\
                          0 & 0 & 1
                        \end{bmatrix}
     \begin{bmatrix} \fden{x} \\ \fnum{y} \\ \fnum{z} \end{bmatrix}
   = \begin{bmatrix} \fden{x} \\ x\fden{x} + z\fnum{y} \\ \fnum{z} \end{bmatrix}
   = \begin{bmatrix} \fden{x} \\ \fnum{x}\\ \fnum{z} \end{bmatrix}
   = \psi_{\rm b}
 \end{equation}
confirms that $\dsp \psi_{\rm b} = \mc{G} \psi_{\rm a}.$

Thus, the representative Lax matrices, \eqref{M:sBoussLaxPairA} and \eqref{M:sBoussLb},
are gauge equivalent, as in \eqref{E:altGauge1}, with $\mc{G}$ in \eqref{M:sB-GaugeAB}.
In essence, for Lax matrices of the same sizes, $\mc{G}$ ``represents" the edge constraint in
the system of \PDeltaEs.
Of course, \eqref{E:altGauge1} may also be represented as
$ L_{\rm a} = \bar{\mc{G}}_1 L_{\rm b} \bar{\mc{G}}^{-1}, $
where $\dsp \bar{\mc{G}} = \mc{G}^{-1}$.

% fixed lower cases 09/16/2019
%%  sub-section Gauge Like Equivalence
\subsection{Gauge-like equivalence}
\label{sec:GaugeLikeEquiv}
 In as much as gauge transformations between the representative Lax matrices of a given system of \PDeltaEs\
 are straightforward to derive and defined by the corresponding edge equation, the relationship between
 representative and extended matrices, though still dependent upon the edge equation, is not so obvious.

 Consider the Schwarzian Boussinesq system \eqref{E:sBouss} which we have shown to have representative Lax matrices,
 $L_{\rm a}$ and $L_{\rm b}$ (derived in Section~\ref{sec:firstAlt1}) and extended Lax matrices,
 $L_{\rm A}$, $L_{\rm B}$, $L_{\rm C},$ and $L_{\rm D}$ derived in Section~\ref{sec:firstAlt2}.

 To determine a relationship between extended Lax matrices (like $L_{\rm A}$) and representative Lax matrices
 (like $L_{\rm a}$), \eqref{E:altGauge1} must be generalized because the matrices do not have the same sizes.
 We therefore introduce transformations involving non-square matrices $\mc{H}$  and $\bar{\mc{H}}$ satisfying
 one of the relationships,
 \begin{subequations}
  \label{E:sB-GaugeLikeEq}
  \begin{align}
   L_{\rm ext} &= \mc{H}_1 L_{\rm rep} \mc{H}_{\lft}^{-1}, \label{E:sB-GaugeR2E} \\
   L_{\rm rep} &= \bar{\mc{H}}_1 L_{\rm ext} \bar{\mc{H}}^{-1}_{\rgt},
  \label{E:sB-GaugeE2R}
  \end{align}
 \end{subequations}
 where $L_{\rm rep}$ is a representative Lax pair, $L_{\rm ext}$ is an extended Lax pair,
 and $\mc{H}$ and $\bar{\mc{H}}$ are suitable matrices of appropriate sizes.
 % fixed Left and Right 09/16/2019
 Furthermore, the labels ``{\rm Left}" and ``{\rm Right}" refer to left and right inverses.
 In deriving $\mc{H}$ and $\bar{\mc{H}}$ we find that the edge equations again provide guidance.

 Obviously, matrices like $\mc{H}$ and $\bar{\mc{H}}$ play the role of the gauge matrices but since they
 are no longer square we call them {\em gauge-like} matrices.
% fixed plural 09/16/2019
 Likewise, any of the transformations in \eqref{E:sB-GaugeLikeEq} are called {\em gauge-like} transformations.
\vskip 8pt
\noindent
{\bf Example 2.}$\;$ To illustrate \eqref{E:sB-GaugeR2E}, consider an extended Lax pair with associated vector $\psi$.
If we consider the edge constraint expressed as in \eqref{E:xyzSubsOpt2}, then the linearity of
$\fnum{y} = -\, \frac{x \fden{x} - \fnum{x}}{z}$ in variables $\fden{x}$ and $\fnum{x}$ allows us to
% fix replace (but not unique) by (unique) 09/16/2019
express $\psi$ in terms of $\psi_{\rm b}$ in a simple (unique) way
 \begin{equation}
  \label{E:sB-GEquivLa2Lay}
  \psi = \begin{bmatrix} \fden{x}\\ \fnum{x}\\ \fnum{y}\\ \fnum{z}\end{bmatrix}
    = \begin{bmatrix}
        \fden{x}\\ \fnum{x}\\
        -\, \frac{x}{z}\fden{x} + \frac{1}{z}\fnum{x}\\
        \fnum{z}\end{bmatrix}
    = \begin{bmatrix} 1&0&0\\ 0&1&0\\ -\frac{x}{z}&\frac{1}{z}&0\\ 0&0&1\end{bmatrix}
    \begin{bmatrix} \fden{x}\\ \fnum{x}\\ \fnum{z}\end{bmatrix}
    := \mc{H} \psi_{\rm b},
 \end{equation}
defining the matrix $\mc{H}.$
Since rank $\mc{H} = 3$, matrix $\mc{H}$ has a 3-parameter family of left inverses,
\begin{equation}
  \mc{H}_{\lft, all}^{-1} =
    \begin{bmatrix}
     1-\alpha x & \alpha & -\alpha z & 0 \\
      (1-\beta) x & \beta & (1-\beta) z & 0 \\
      -\gamma x & \gamma & -\gamma z & 1
    \end{bmatrix},
 \end{equation}
 where $\alpha$, $\beta$ and $\gamma$ are free parameters (which could depend on ${\bf{x}}$).

Now we take a specific member of the family (denoted by $\mc{H}_{\lft}^{-1})$ so that
\begin{equation}
\label{E:G-B2AxyNew1}
L_{\rm C} \;\;\dot{=}\;\; \mc{H}_1 L_{\rm b} \mc{H}_{\lft}^{-1},
\end{equation}
where, as before, $\dot{=}$ indicates equality when evaluated against the given \PDeltaEs.
More precisely, equality only holds when edge equation \eqref{E:sBoussflooralla} is used.
A straightforward matrix multiplication shows that \eqref{E:G-B2AxyNew1} holds if
$\alpha = \beta = \gamma = 0$.
Hence,
 \begin{equation}
  \label{E:G-B2AxyNew2}
   \mc{H}_{\lft}^{-1} =
   \begin{bmatrix} 1&0&0&0\\ x&0&z&0\\ 0&0&0&1\end{bmatrix}.
 \end{equation}
% fixed E:G-B2AxyNew1 instead of E:G-B2AxyNew2  09/16/2019
Instead of \eqref{E:altGauge1} we now have \eqref{E:G-B2AxyNew1},
i.e., a transformation of type \eqref{E:sB-GaugeR2E}, which can readily be verified.
% fixed repeatedly was in wrong place 09/16/2019
Indeed, repeatedly using \eqref{E:G-B2AxyNew1}, \eqref{E:sB-GEquivLa2Lay}, and \eqref{LP:matrix}, yields
\begin{equation}
\label{E:altGaugeAC}
\psi_1
  = L_{\rm C} \psi
  = \mc{H}_1 L_{\rm b} \mc{H}^{-1}_{\rm Left} \mc{H} \psi_{\rm b}
  = \mc{H}_1 L_{\rm b} \psi_{\rm b}
  = \mc{H}_1 \left( \psi_{\rm b} \right)_1
  = \Big( \mc{H} \psi_{\rm b} \Big)_1,
\end{equation}
confirming \eqref{E:sB-GEquivLa2Lay}.
\vskip 8pt
\noindent
{\bf Example 3.}$\;$
After similar calculations involving
$\psi_{\rm a} = \begin{bmatrix} \fden{x} & \fnum{y} & \fnum{z} \end{bmatrix}^{\rm T}$ and $\psi$,
and with the edge constraint expressed as in
\eqref{E:xyzSubsx3}, i.e., $\fnum{x} = x \fden{x} + z \fnum{y}$, we find that
 \begin{equation}
  \label{M:G-A2Bz}
   L_{\rm D} \;\;\dot{=}\;\; \mc{H}_1 L_{\rm a} \mc{H}_{\lft}^{-1},
 \end{equation}
 where
 \begin{equation}
  \label{M:G-A2BzH}
   \mc{H} = \begin{bmatrix} 1&0&0\\ x&z&0\\ 0&1&0\\ 0&0&1\end{bmatrix}
   \;\;\text{ and }\;\;
   \mc{H}_{\lft}^{-1} =
         \begin{bmatrix} 1&0&0&0\\ -x/z&1/z&0&0\\ 0&0&0&1\end{bmatrix}.
 \end{equation}
\vskip 8pt
\noindent
{\bf Example 4.}$\;$  A first gauge-like relationship between $L_{\rm a}$ and $L_{\rm A}$ is simple to derive.
As mentioned in Choice 1 in Section~\ref{sec:firstAlt2}, removing the second row and second column from
$L_{\rm A}$ gives $L_{\rm a}$.
Formally,
\begin{equation}
\label{E:sB-gaugeP2Pa-1bis}
    L_{\rm a} \;\;\dot{=}\;\;
    \mc{B} L_{\rm A} \mc{B}^{\rm T},
 \end{equation}
with $z y_1 = x_1 - x$ and
\begin{equation}
\label{E:matB}
\mc{B} = \begin{bmatrix} 1&0&0&0\\ 0&0&1&0\\ 0&0&0&1 \end{bmatrix}.
\end{equation}
Continuing with $L_{\rm a}$ and $L_{\rm A}$, we derive a second gauge-like transformation
to illustrate \eqref{E:sB-GaugeE2R}.
To find a matrix $\bar{\mc{H}}$, consider the edge constraint $\fnum{y} = -\, \frac{x \fden{x} - \fnum{x}}{z}$
expressed in \eqref{E:xyzSubsOpt2}.
Thus,
 \begin{equation}
  \label{E:sB-gaugeP2Pa-2}
   \psi_{\rm a} = \begin{bmatrix} \fden{x}\\ \fnum{y}\\ \fnum{z}\end{bmatrix}
          = \begin{bmatrix} \fden{x}\\ -\frac{x}{z}\fden{x} + \frac{1}{z}\fnum{x} \\ \fnum{z}\end{bmatrix}
          = \begin{bmatrix}
             1               & 0           & 0 & 0\\
             -\,\frac{x}{z} & \frac{1}{z} & 0 & 0\\
             0               & 0           & 0 & 1 \\
            \end{bmatrix}
            \begin{bmatrix} \fden{x}\\ \fnum{x}\\ \fnum{y}\\ \fnum{z}\end{bmatrix}
          := \bar{\mc{H}}\psi.
 \end{equation}
 % fixed = added and G replaced by g  09/16/2019
 The inverse transformation,
 \begin{equation}
  \label{E:sB-gaugePa2P}
   \psi = \begin{bmatrix} \fden{x}\\ \fnum{x}\\ \fnum{y}\\ \fnum{z}\end{bmatrix}
         = \begin{bmatrix} \fden{x} \\ x \fden{x} + z \fnum{y}\\ \fnum{y}\\ \fnum{z} \end{bmatrix}
          = \begin{bmatrix}
             1 & 0 & 0 \\
             x & z & 0 \\
             0 & 1 & 0\\
             0 & 0 & 1
            \end{bmatrix}
            \begin{bmatrix} \fden{x}\\ \fnum{y}\\ \fnum{z}\end{bmatrix}
          := \bar{\mc{H}}_{\rgt}^{-1} \psi_{\rm a},
 \end{equation}
determines a suitable right inverse of $\bar{\mc{H}}.$
Thus,
 \begin{equation}
  \label{M:sB-GaugeEquivCB}
    L_{\rm a} = \bar{\mc{H}}_1 L_{\rm A} \bar{\mc{H}}^{-1}_{\rgt}  \;\;\text{ for }\;\;
     \bar{\mc{H}} = \begin{bmatrix}
               1&0&0&0\\
               -\, \frac{x}{z} & \frac{1}{z} &0&0\\
               0&0&0&1
              \end{bmatrix},
 \end{equation}
without having to use the edge equation \eqref{E:sBoussflooralla}.
To show that \eqref{M:sB-GaugeEquivCB} is correct, use \eqref{LP:matrix} repeatedly, together with
\eqref{E:sB-gaugePa2P} and \eqref{M:sB-GaugeEquivCB}, yielding
\begin{equation}
\label{E:altGaugeACalter}
\left( \psi_{\rm a} \right)_1
  = L_{\rm a} \psi_{\rm a}
  = {\bar{\mc{H}}}_1 L_{\rm A} {\bar{\mc{H}}}^{-1}_{\rm Right} \psi_{\rm a}
  = {\bar{\mc{H}}}_1 L_{\rm A} \psi
  = {\bar{\mc{H}}}_1 \psi_1
  = \Big( {\bar{\mc{H}}} \psi \Big)_1,
\end{equation}
confirming \eqref{E:sB-gaugeP2Pa-2}.
\vskip 8pt
\noindent
{\bf Example 5.}$\;$ Interestingly, gauge transformations between two distinct extended Lax matrices for a \PDeltaE\
are not as straightforward.
For example, a gauge transformation has not yet been found for Lax matrices $L_{\rm A}$ and $L_{\rm B}$ given in
\eqref{M:sBoussLc} and \eqref{M:sBoussLd}.
Thus, even though we have shown that the corresponding representative Lax matrices are gauge equivalent,
we have not been able to show the same for the corresponding extended matrices.

% fixed lower cases 09/16/2019
% SECTION 5 -- APPLICATION
\section{Application to generalized Hietarinta systems}
\label{sec:sBoussApplic}

% fixed presented the results of a search, edge-constrained 09/16/2019
In \cite{ref:Hietarinta2011}, Hietarinta presented the results of a search of multi-component equations
which are edge-constrained and obey the property of multidimensional consistency.
% fixed past tense led instead of lead 09/16/2019
That search led to various generalized Boussinesq-type systems, nowadays called the Hietarinta
A-2, B-2, C-3, and C-4 systems.
Bridgman {\em et al.} \cite{ref:Bridgmanetal2013} derived their corresponding Lax pairs using \cite{ref:Bridgman2015}.

% fixed added the, replaced subsequently by simultaneously 09/16/2019
Simultaneously, Zhang {\em et al.} \cite{ref:Zhangetal2012} showed that each of the lattice systems presented
in \cite{ref:Hietarinta2011} can be further generalized based on a direct linearization scheme
\cite{ref:Nijhoffetal1992} in connection with a more general dispersion law.
The systems considered in \cite{ref:Hietarinta2011} are then shown to be special cases.
In fact, they are connected to the more general cases through point transformations.

As a by-product of the direct linearization method, Zhang {\em et al.} \cite{ref:Zhangetal2012}
obtained the Lax pairs of each of these generalized Boussinesq systems.
No doubt, they are all valid Lax pairs but some of the matrices have larger than needed sizes.
Using the algorithmic CAC approach discussed in Section~\ref{sec:Boussinesqsystems},
we were able to derive Lax pairs of minimal matrix sizes for these systems and unravel
the connections with the Lax matrices presented in \cite{ref:Zhangetal2012}.
Full details of that investigation will be published elsewhere \cite{ref:BridgmanHereman2018}
but their Lax matrices are given in the next section.

% fixed lower cases 09/16/2019
% SECTION 6 -- SUMMARY RESULTS
\section{Summary of results}
\label{sec:sBoussSummary}
%  fixed lower cases 09/16/2019
\subsection{Lattice Boussinesq system}
\label{sec:BoussGaugeEquivalence}
 The lattice Boussinesq system \cite{ref:Tongas2005} is given by
 \begin{equation}
  \label{E:Bouss}
  \begin{aligned}
   & z_1 - xx_1 + y = 0, \;\;
     z_2 - xx_2 + y = 0, \; \text{ and } \\
   & (x_2 - x_1)(z - xx_{12} + y_{12}) - p + q = 0.
  \end{aligned}
 \end{equation}
 Edge constraint $\dsp x_1 = \frac{z_1 + y}{x}$ implies that $\dsp x_3 = \frac{z_3 + y}{x}$.
 Here,
 \begin{equation}
  \label{E:Bouss13cc}
  \begin{aligned}
   x_{13} &= \frac{y_1 - y_3}{x_1 - x_3}, \;\;
   y_{13} = \frac{x(y_1 - y_3) - z(x_1 - x_3) + k - p}{x_1 - x_3}, \;\; \text{ and } \\
   z_{13} &= \frac{x_3y_1 - x_1y_3}{x_1 - x_3}.
  \end{aligned}
 \end{equation}
 Variants of \eqref{E:Bouss13cc} may be derived by incorporating edge constraints yielding
 \begin{equation}
  \label{E:Bouss13A}
  \begin{aligned}
   \tilde{x}_{13} &= \frac{x(y_1 - y_3)}{z_1 - z_3}, \;\;
   \tilde{y}_{13} = \frac{x\Big(k - p + x(y_1 - y_3)\Big) - z(z_1 - z_3)}{z_1 - z_3}, \; \text{ and } \\
   \tilde{z}_{13} &= \frac{y(y_1 - y_3) + (y_1z_3 - y_3z_1)}{z_1 - z_3}.
  \end{aligned}
 \end{equation}
Using the algorithm of Section~\ref{sec:Boussinesqsystems}, we computed $3$ by $3$ Lax matrices which are
presented in Table 1 together with the gauge transformations.
For $L_{\rm a}$ one has $\frac{s \, t_2}{t \, s_1} \;\dot{=}\; 1,$ which is satisfied for $t = s = 1.$
For $L_{\rm b}$ one obtains $\frac{s \, t_2}{t \, s_1} \;\dot{=}\; \frac{x_1}{x_2},$ hence, $t = s = \frac{1}{x}.$

Using only \eqref{E:Bouss13cc} with \eqref{E:xyzSubs} yields a $4$ by $4$ Lax matrix (not shown) that is
trivially associated with $L_{\rm a}$.
Similarly, using only \eqref{E:Bouss13A} again with \eqref{E:xyzSubs} gives a $4$ by $4$ Lax matrix (not shown)
which is trivially associated with $L_{\rm b}$.
\vfill
\newpage
% fixed lower cases in Table caption 09/16/2019
% TABLE 1
\begin{longtable}{lcl}
%  \caption{Boussinesq System Lax Pairs \& Gauge Matrices}
\caption{Boussinesq system Lax pairs and gauge matrices}
 \label{T:Bouss}\\
 \toprule[1pt]
  Substitutions & $\psi$ & Matrices $L$ of Lax pair \\
 \midrule
 \multicolumn{3}{l}{Writing the edge constraint as $\dsp z_3 = xx_3 - y$ yields}\\
 \addlinespace[1em]
  \parbox{4.5cm}{
   $\begin{aligned}
   	x_3 &= \frac{f}{F},\ \
   	y_3 = \frac{g}{F}, \\
   	z_3 &= -\frac{y F - x f}{F}.
   \end{aligned}$
  } &
  \parbox{2.5cm}{
   $\psi_{\rm a} = \begin{bmatrix} F\\ f\\ g\end{bmatrix}$
  } &
  \parbox{5.5cm}{
  $\hfill \dsp L_{\rm a} = \begin{bmatrix}
          -x_1 & 1 & 0\\
          -y_1 & 0 & 1\\
          \ell_{31}  & -z & x
         \end{bmatrix}$,}\\
 \addlinespace[1em]
 \multicolumn{3}{l}{
  where $\ell_{31} = zx_1 - xy_1 + p - k$.} \\
 \addlinespace[0.9em]
 \midrule
 \multicolumn{3}{l}{Writing the edge constraint as $x_3 = \frac{z_3 + y}{x}$ yields}\\
  \parbox{4.5cm}{
   $\begin{aligned}
   	x_3 &= \frac{y F + h }{xF}, \\
   	y_3 &= \frac{g}{F},\ \
   	z_3 = \frac{h}{F}.
   \end{aligned}$
  } &
  \parbox{2.5cm}{
   $\psi_{\rm b} = \begin{bmatrix} F\\ g\\ h\end{bmatrix}$
  } &
  \parbox{5.5cm}{
  $\hfill \dsp L_{\rm b} = \frac{1}{x}\, \begin{bmatrix}
          y - xx_1 & 0 & 1\\
          \ell_{21} & x^2 & -z\\
           -yy_1 & xx_1 & -y_1
         \end{bmatrix}$,
         } \\
 \addlinespace[1em]
 \multicolumn{3}{l}{
 where $\ell_{21} = x(p - k - xy_1) - z(y - xx_1)$.
 } \\
 \addlinespace[0.9em]
 \midrule
 \multicolumn{3}{l}{Gauge transformations for $L_{\rm a}$ and $L_{\rm b}$ are given by }\\
 \addlinespace[1em]
 \multicolumn{3}{l}{\parbox{12.5cm}{
   \begin{tabular}{p{5.5cm} p{3.5cm} p{3.5cm}}
    $\begin{aligned}
     L_{\rm b} &= \mc{G}_1 L_{\rm a} \mc{G}^{-1}, \psi_{\rm b} = \mc{G} \psi_{\rm a}, \\
     L_{\rm a} &= \bar{\mc{G}}_1 L_{\rm b} \bar{\mc{G}}^{-1},  \psi_{\rm a} = \bar{\mc{G}} \psi_{\rm b},
    \end{aligned}$
    &
    $\mc{G} = \begin{bmatrix} 1 & 0 & 0 \\ 0 & 0 & 1 \\ -y & x & 0 \end{bmatrix} $
    & \hfill
    $\bar{\mc{G}} = \begin{bmatrix} 1 & 0 & 0 \\ \frac{y}{x} & 0 & \frac{1}{x} \\ 0 & 1 & 0 \end{bmatrix}. $ \\
    \multicolumn{3}{l}{where $\bar{\mc{G}} = \mc{G}^{-1}$.}
   \end{tabular}
  }
 } \\
 \addlinespace[0.8em]
 \bottomrule[1pt]
\end{longtable}
\vspace*{1mm}
\noindent
%%%  Schwarzian Boussinesq System
% fixed lower cases 09/16/2019
\subsection{Schwarzian Boussinesq system}
\label{sec:SBoussGaugeEquivalence}
The Schwarzian Boussinesq system \cite{ref:Hietarinta2011,ref:Nijhoff1997,ref:Nijhoff1999} is given by
 \begin{equation}
  \label{E:SBouss}
  \begin{aligned}
   &z y_1 - x_1 + x = 0,\;\;
    z y_2 - x_2 + x = 0, \;\; \text{ and } \\
   & (z_1 - z_2) z_{12} - \frac{z}{y}\left(p y_1 z_2 - q y_2 z_1\right) = 0.
  \end{aligned}
 \end{equation}
 Edge constraint $\dsp x_1 = z y_1 + x$ leads to $\dsp x_3 = z y_3 + x$.
 Hence,
 \begin{equation}
  \label{E:SBouss13cc}
  \begin{aligned}
   x_{13} &= \frac{x_3z_1 - x_1z_3}{z_1 - z_3}, \;\;
     y_{13} = \frac{x_3 - x_1}{z_1 - z_3}, \; \text{ and } \\
   z_{13} &= \frac{z}{y}\left(\frac{p y_1z_3 - k y_3z_1}{z_1 - z_3}\right).
  \end{aligned}
 \end{equation}
 After incorporating edge constraints, variants of \eqref{E:SBouss13cc} are
 \begin{equation}
  \label{E:SBouss13sA}
  \begin{aligned}
   \tilde{x}_{13}   &= \frac{x(z_1 - z_3) + z(y_3z_1 - y_1z_3)}{z_1 - z_3}, \;\;
   \tilde{y}_{13}    = \frac{z(y_3 - y_1)}{z_1 - z_3}, \;\; \text{ and } \\
     \tilde{z}_{13} &= \frac{k z_1(x - x_3) - p z_3(x - x_1)}{y(z_1 - z_3)}.
  \end{aligned}
 \end{equation}
 The $3$ by $3$ matrices computed in Section~\ref{sec:Boussinesqsystems} are summarized in Table 2
 together with the gauge transformations that connect them.
 For the representative and extended Lax matrices given below we obtained
 $\frac{s \, t_2}{t \, s_1} \; \dot{=}\; \frac{z_1}{z_2},$ which holds when $t = s = \frac{1}{z}.$
% fixed lower cases in Table caption 09/16/2019
% TABLE 2
\begin{longtable}{lcl}
% \caption{Schwarzian Boussinesq System Lax Pairs \& Gauge Matrices}
\caption{Schwarzian Boussinesq system Lax pairs and gauge matrices}
 \label{T:SBouss}\\
 \toprule[1pt]
  Substitutions & $\psi$ & Matrices $L$ of Lax pair \\
 \midrule
 \multicolumn{3}{l}{Writing the edge constraint as $x_3 = zy_3 + x$ yields}\\
 \addlinespace[1em]
  \parbox{4cm}{
   $\begin{aligned}
   	x_3 &= \frac{xF + zg}{F}, \\
   	y_3 &= \frac{g}{F}, \ \
   	z_3 = \frac{h}{F}.
   \end{aligned}$
  } &
  \parbox{2.5cm}{
   $\psi_{\rm a} = \begin{bmatrix} F\\ g\\ h\end{bmatrix}$
  } &
  \parbox{6cm}{ \hfill
  $\dsp L_{\rm a} = \frac{1}{z}\begin{bmatrix}
          z_1 & 0 & -1\\
          x - x_1 & z & 0\\
          0  & -\,\frac{k zz_1}{y} & \frac{p zy_1}{y}
         \end{bmatrix}$.} \\
 \addlinespace[1em]
 \midrule
 \multicolumn{3}{l}{Writing the edge constraint as $y_3 = \frac{x_3 - x}{z}$ yields}\\
 \addlinespace[1em]
  \parbox{4cm}{
   $\begin{aligned}
   	x_3 &= \frac{f}{F},\ \  z_3 = \frac{h}{F}, \\
   	y_3 &= -\frac{xF - f}{zF}.
   \end{aligned}$
  } &
  \parbox{2.5cm}{
   $\psi_{\rm b} = \begin{bmatrix} F\\ f\\ h\end{bmatrix}$
  } &
  \parbox{6cm}{\hfill
  $\dsp L_{\rm b} = \frac{1}{z}\begin{bmatrix}
          z_1 & 0 & -1\\
          0 & z_1 & -x_1\\
          \frac{k xz_1}{y} & -\,\frac{k z_1}{y} & \frac{p zy_1}{y}
         \end{bmatrix}$.} \\
 \addlinespace[1em]
 \midrule
 \multicolumn{3}{l}{Gauge transformations for $L_{\rm a}$ and $L_{\rm b}$ are given by}\\
 \multicolumn{3}{l}{\parbox{12.5cm}{
   \begin{tabular}{p{5.5cm} p{3.5cm} p{3.5cm}}
    $\begin{aligned}
     L_{\rm b} &= \mc{G}_1 L_{\rm a} \mc{G}^{-1}, \psi_{\rm b} = \mc{G} \psi_{\rm a}, \\
     L_{\rm a} &= \bar{\mc{G}}_1 L_{\rm b} \bar{\mc{G}}^{-1},  \psi_{\rm a} = \bar{\mc{G}} \psi_{\rm b},
    \end{aligned}$
    &
    $\mc{G} = \begin{bmatrix} 1 & 0 & 0 \\ x & z & 0 \\ 0 & 0 & 1 \end{bmatrix} $
    & \hfill
    $\bar{\mc{G}} = \begin{bmatrix} 1 & 0 & 0 \\ -\,\frac{x}{z} & \frac{1}{z} & 0 \\ 0 & 0 & 1 \end{bmatrix}, $ \\
    \multicolumn{3}{l}{where $\bar{\mc{G}} = \mc{G}^{-1}$.}
   \end{tabular}
  }
 } \\
 \addlinespace[1em]
 \bottomrule[1pt]
\end{longtable}
\vskip 3pt
\noindent
System \eqref{E:SBouss} also admits the extended Lax matrices,
\begin{equation}
% \label{M:sBoussLy}
\!\!\!\!L_{\rm A} \!=\! \frac{1}{z}\!
  \begin{bmatrix}\!
   z_1 & 0 & 0 & -1\\
   0 & z_1 & 0 & -x_1\\
   -zy_1 & 0 & z & 0\\
   0 & 0 & -\frac{k zz_1}{y} & \frac{p zy_1}{y}\!\!
  \end{bmatrix}
  \,\text{ and }\,
 L_{\rm B} \!=\! \frac{1}{z}\!
   \begin{bmatrix}\!
    z_1 & 0 & 0 & -1\\
    0 & z_1 & 0 & -x_1\\
    -zy_1 & 0 & z & 0\\
    \frac{k xz_1}{y} & -\frac{k z_1}{y} & 0 & -\frac{p (x-x_1)}{y}\!\!
   \end{bmatrix}\!\!,
\end{equation}
when considering the edge-modified forms of $y_{13}$, and of $y_{13}$ and $z_{13}$, respectively;
and
\begin{equation}
\!\!\!\!L_{\rm C} \!=\! \frac{1}{z}\!\!
  \begin{bmatrix}\!
   z_1 & 0 & 0 & -1\\
   xz_1 & 0 & zz_1 & -(x + z y_1)\!\\
   -zy_1 & 0 & z & 0\\
   0 & 0 & -\frac{k zz_1}{y}  & \frac{p zy_1}{y}\!\!
  \end{bmatrix}\!
  \text{ and }
 L_{\rm D} \!=\! \frac{1}{z}\!\!
   \begin{bmatrix}\!
    z_1 & 0 & 0 & -1\\
    0 & z_1 & 0 & -x_1\\
    -x_1 & 1 & 0 & 0\\
    \frac{k xz_1}{y} & -\frac{k z_1}{y} & 0 & \frac{p (x_1 - x)}{y}\!\!
\end{bmatrix}\!\!,
\end{equation}
when considering the edge-modified forms of $x_{13}$ and $y_{13}$, and of $z_{13}$, respectively.
All other combinations of \eqref{E:SBouss13cc} and \eqref{E:SBouss13sA} result in matrices
which do not satisfy the defining equation \eqref{matLaxEqdiffeqs}.
% fixed lower cases 09/16/2019
\subsection{Generalized Hietarinta systems}
\label{sec:genHietsys}
In \cite{ref:Zhangetal2012}, the authors introduced generalizations of Hietarinta's systems \cite{ref:Hietarinta2011}
by considering a general dispersion law,
 \begin{equation}
  \label{E:disprel}
	\nG(\omega, \kappa) := \omega^3 - \kappa^3 + \alpha_2(\omega^2 - \kappa^2) + \alpha_1(\omega - \kappa),
 \end{equation}
where $\alpha_1$ and $\alpha_2$ are constant parameters.
For example, for the special case $a = \alpha_1 = \alpha_2 = 0,$ one gets $\nG(-p, -a) = -p^3$ and $\nG(-q, -a) = -q^3.$
Then \eqref{E:gHA2} (below) reduces to Hietarinta's original A-2 system in \cite[p.\ 95]{ref:HietarintaJoshiNijhoff2016}.
The term with coefficient $b_0$ could be removed by a simple transformation \cite{ref:Hietarinta2011}.
We will keep it to cover the most general case.
In \cite{ref:Bridgmanetal2013,ref:Hietarinta2011}, $p^3$ and $q^3$ are identified with $p$ and $q,$ respectively.

The explicit form of $\nG(\omega, \kappa)$ in \eqref{E:disprel} is not needed \cite{ref:BridgmanHereman2018}
to compute Lax pairs.
However, for the B-2 system the condition
\begin{equation}
 \label{E:dispRelConstB2}
 \nG(-p, -k) + \nG(-k, -q) = \nG(-p, -q)
\end{equation}
{\it must hold} for 3D consistency and, consequently, for the computation of Lax pairs.

Zhang {\em et al.} \cite{ref:Zhangetal2012} computed $4$ by $4$ Lax matrices for these generalized systems
with the direct linearization method.
By incorporating the edge equations (as shown in Section~\ref{sec:Boussinesqsystems}),
we were able to find $3$ by $3$ matrices which are presented in this section.
Computational details will appear in a forthcoming paper \cite{ref:BridgmanHereman2018}.

%%%  Generalized Hietarinta A-2 System
% fixed lower cases 09/16/2019
\subsubsection{Generalized Hietarinta A-2 system}
\label{sec:gHA2GaugeEquivalence}
 The generalized Hietarinta A-2 system \cite{ref:Zhangetal2012} is given by
 \begin{equation}
  \label{E:gHA2}
  \begin{aligned}
   &z x_1 - y_1 - x = 0, \;\;  z x_2 - y_2 - x = 0, \;\;\text{ and } \\
   &y - x z_{12} + b_0 x + \frac{\nG(-p,-a) x_1 - \nG(-q, -a) x_2}{z_2 - z_1} = 0.
  \end{aligned}
 \end{equation}
 From edge constraint $\dsp x_1 = \frac{x + y_1}{z}$ one gets $\dsp x_3 = \frac{x + y_3}{z}$.
 Here,
 \begin{equation}
  \label{E:gHA213cc}
  \begin{aligned}
   x_{13} &= \frac{x_1 - x_3}{z_1 - z_3}, \;\; y_{13} = \frac{x_1 z_3 - x_3 z_1}{z_1 - z_3}, \;\; \text{ and} \\
   z_{13} &= \frac{(y + b_0 x) (z_1 - z_3) + \nG(-k,-a) x_3 - \nG(-p,-a) x_1}{x (z_1 - z_3)}.
  \end{aligned}
 \end{equation}
The $3$ by $3$ Lax matrices are presented in Table 3, together with the gauge transformations
that connect them.
For $L_{\rm a}$ one has $\frac{s \, t_2}{t \, s_1} \; \dot{=}\; \frac{z_1}{z_2},$ hence, $t = s = \frac{1}{z}$.
For $L_{\rm b}$ we set $t = s = 1$ since $\frac{s \, t_2}{t \, s_1} \; \dot{=}\; 1$.

Alternative forms of \eqref{E:gHA213cc} (after incorporating edge constraints) are
 \begin{equation}
  \label{E:gHA213sA}
  \begin{aligned}
   \tilde{x}_{13} &= \frac{y_1 - y_3}{z(z_1 - z_3)}, \;\;
   \tilde{y}_{13} = - \left( \frac{x}{z} + \frac{y_3 z_1 - y_1 z_3}{z(z_1 - z_3)} \right), \;\; \text{ and} \\
   \tilde{z}_{13} &= \frac{y + b_0x}{x} + \frac{ \nG(-k,-a)(x + y_3) - \nG(-p,-a)(x + y_1)}{xz(z_1 - z_3)}.
  \end{aligned}
 \end{equation}
Using only \eqref{E:gHA213cc} with \eqref{E:xyzSubs} leads to a $4$ by $4$ Lax matrix (not shown)
that is trivially associated with $L_{\rm a}$.
Similarly, as shown in \cite{ref:Zhangetal2012}, using only \eqref{E:gHA213sA} with \eqref{E:xyzSubs} results
in a $4$ by $4$ Lax matrix (not shown) which is trivially associated with $L_{\rm b}$.
\vfill
\newpage
% fixed lower cases in Table caption 09/16/2019
% TABLE 3
\begin{longtable}{lcl}
 % \caption{Gen. Hietarinta A-2 System Lax Pairs \& Gauge Matrices}
 \caption{$\!$Generalized Hietarinta A-2 system Lax pairs and gauge matrices$\!$}
 \label{T:gHA2}\\
 \toprule[1pt]
  Substitutions & $\psi$ & Matrices $L$ of Lax pair \\
 \midrule
 \multicolumn{3}{l}{Writing the edge constraint as $x_3 = \frac{y_3 + x}{z}$ yields} \\
  \parbox{3.5cm}{
   $\begin{aligned}
   	x_3 &= \frac{xF + g}{zF}, \\
   	y_3 &= \frac{g}{F},\ \
   	z_3 = \frac{h}{F}.
   \end{aligned}$
  } &
  \parbox{2.5cm}{
   $\psi_{\rm a} = \begin{bmatrix} F\\ g\\ h\end{bmatrix}$
  } &
  \parbox{6.5cm}{
  $\dsp L_{\rm a} = \frac{1}{z}\begin{bmatrix}
          -zz_1 & 0 & z \\
          xz_1 & z_1 & -zx_1 \\
          \ell_{31}  & -\,\frac{\nG(-k,-a)}{x} & \frac{(y + b_0x)z}{x}
         \end{bmatrix}$,\\
        } \\
 \addlinespace[1em]
 \multicolumn{3}{l}{where $\ell_{31} = \frac{1}{x}\big(\nG(-p,-a)zx_1 - \nG(-k,-a)x - (y + b_0x)zz_1\big)$.}\\
 \addlinespace[0.70em]
 \midrule
 \multicolumn{3}{l}{Writing the edge constraint as $\dsp y_3 = zx_3 - x$ yields}\\
  \parbox{3.5cm}{
   $\begin{aligned}
   	x_3 &= \frac{f}{F},\ \ z_3 = \frac{h}{F},  \\
   	  y_3 &= -\frac{xF - zf}{F}.
   \end{aligned}$
  } &
  \parbox{2.5cm}{
   $\psi_{\rm b} = \begin{bmatrix} F\\ f\\ h\end{bmatrix}$
  } &
  \parbox{6.5cm}{ \hfill
  $\dsp L_{\rm b} = \begin{bmatrix}
          - z_1 & 0 & 1\\
          - x_1 & 1 & 0\\
          \tilde{\ell}_{31} & -\,\frac{\nG(-k, -a)}{x} & \frac{y + b_0x}{x}
         \end{bmatrix}$,} \\
 \addlinespace[1em]
 \multicolumn{3}{l}{where $\tilde{\ell}_{31} = \frac{1}{x}\big(\nG(-p,-a)x_1 - (y + b_0x)z_1\big)$.}\\
 \addlinespace[0.70em]
 \midrule
 \multicolumn{3}{l}{Gauge transformations for $L_{\rm a}$ and $L_{\rm b}$ are given by}\\
 \multicolumn{3}{l}{\parbox{12.5cm}{
   \begin{tabular}{p{5.5cm} p{3.5cm} p{3.5cm}}
    $\begin{aligned}
     L_{\rm b} &= \mc{G}_1 L_{\rm a} \mc{G}^{-1}, \psi_{\rm b} = \mc{G} \psi_{\rm a}, \\
     L_{\rm a} &= \bar{\mc{G}}_1 L_{\rm b} \bar{\mc{G}}^{-1},  \psi_{\rm a} = \bar{\mc{G}} \psi_{\rm b},
    \end{aligned}$
    &
    $\mc{G} = \begin{bmatrix} 1 & 0 & 0 \\ \frac{x}{z} & \frac{1}{z} & 0 \\ 0 & 0 & 1 \end{bmatrix} $
    & \hfill
    $\bar{\mc{G}} = \begin{bmatrix} 1 & 0 & 0 \\ -x & z & 0 \\ 0 & 0 & 1 \end{bmatrix}, $\\
    \multicolumn{3}{l}{where $\bar{\mc{G}} = \mc{G}^{-1}$.}
   \end{tabular}
  }
 } \\
 \addlinespace[0.450em]
 \bottomrule[1pt]
\end{longtable}
\vspace*{-4mm}
\noindent
%%%  Generalized Hietarinta B-2 System
% fixed lower cases 09/16/2019
\subsubsection{Generalized Hietarinta B-2 system}
\label{sec:gHB2GaugeEquivalence}
 The generalized Hietarinta B-2 system \cite{ref:Zhangetal2012},
 \begin{equation}
  \label{E:gHB2}
  \begin{aligned}
   &x x_1 - z_1 - y = 0, \;\;
     x x_2 - z_2 - y = 0, \;\; \text{ and } \\
   &y_{12} + \alpha_1 + z + \alpha_2(x_{12} - x) - x x_{12} + \frac{\nG(-p,-q)}{x_2 - x_1} = 0,
  \end{aligned}
 \end{equation}
 has edge constraint $\dsp x_1 = \frac{z_1 + y}{x}$ which yields $\dsp x_3 = \frac{z_3 + y}{x}.$
 Here,
 \begin{equation}
  \label{E:gHB213cc}
  \begin{aligned}
  x_{13} &= \frac{y_1 - y_3}{x_1 - x_3}, \;\; z_{13} = \frac{x_3y_1 - x_1y_3}{x_1 - x_3}, \;\; \text{ and } \\
  y_{13} &= (\alpha_2x - \alpha_1 - z) + \frac{(x - \alpha_2)(y_1 - y_3) + \nG(-p,-k)}{x_1 - x_3}.
  \end{aligned}
 \end{equation}
The $3$ by $3$ Lax matrices with their gauge transformations are listed in Table 4.
For $L_{\rm a}$ one has $\frac{s \, t_2}{t \, s_1} \; \dot{=}\; \frac{x_1}{x_2},$ hence, $t = s = \frac{1}{x}$.
For $L_{\rm b}$ we set $t = s = 1$ since $\frac{s \, t_2}{t \, s_1} \; \dot{=}\; 1$.

Other forms of \eqref{E:gHB213cc} by incorporating edge constraints are
 \begin{equation}
  \label{E:gHB213sA}
  \begin{aligned}
   \tilde{x}_{13} &= \frac{x(y_1 - y_3)}{z_1 - z_3}, \;\; \tilde{z}_{13} = -\, \frac{y(y_3 - y_1) + y_3 z_1 - y_1 z_3}{z_1 - z_3},
   \;\; \text{ and }  \\
   \tilde{y}_{13} &= (\alpha_2x - \alpha_1 - z) +
        \frac{x\Big((x - \alpha_2)(y_1 - y_3) + \nG(-p,-k)\Big)}{z_1 - z_3}.
  \end{aligned}
 \end{equation}
  As shown in \cite{ref:Zhangetal2012}, using only \eqref{E:gHB213sA} with \eqref{E:xyzSubs} leads to a $4$ by $4$ Lax matrix
  (not shown) that is trivially associated with $L_{\rm b}$.
Using only \eqref{E:gHB213cc} with \eqref{E:xyzSubs} results in a $4$ by $4$ Lax matrix (not shown) which is
trivially associated with $L_{\rm a}$ when evaluated against the given system.
% fixed lower cases in Table caption 09/16/2019
% TABLE 4
\begin{longtable}{lcl}
 % \caption{Gen. Hietarinta B-2 System Lax Pairs \& Gauge Matrices}
 \caption{$\!$Generalized Hietarinta B-2 system Lax pairs and gauge matrices$\!$}
 \label{T:gHB2}\\
 \toprule[1pt]
  Substitutions & $\psi$ & Matrices $L$ of Lax pair \\
 \midrule
 \multicolumn{3}{l}{Writing the edge constraint as $x_3 = \frac{y + z_3}{x}$ yields}\\
 \addlinespace[1em]
  \parbox{4cm}{
   $\begin{aligned}
   	x_3 &= \frac{yF + h}{xF}, \\
   	y_3 &= \frac{g}{F}, \ \
   	z_3 = \frac{h}{F}.
   \end{aligned}$
  } &
  \parbox{2cm}{
   $\psi_{\rm a} = \begin{bmatrix} F\\ g\\ h\end{bmatrix}$
  } &
  \parbox{6.5cm}{
  $\dsp L_{\rm a} = \frac{1}{x}\begin{bmatrix}
          y - xx_1 & 0 & 1\\
          \ell_{21} & x(x - \alpha_2) & \ell_{23}\\
          -yy_1  & xx_1 & -y_1
         \end{bmatrix}$,} \\
 \addlinespace[0.8em]
 \multicolumn{3}{l}{
  \parbox{12cm}{where $\ell_{21} = (\alpha_2x - \alpha_1 - z)(y - xx_1) + x\big( (\alpha_2 - x)y_1 - \nG(-p,-k) \big)$
   and $\ell_{23} = \alpha_2x - \alpha_1 - z$.}
  }\\
 \addlinespace[1em]
 \midrule
 \multicolumn{3}{l}{Writing the edge constraint as $z_3 = xx_3 - y$ yields}\\
 \addlinespace[1em]
  \parbox{4cm}{
   $\begin{aligned}
   	x_3 &= \frac{f}{F}, \ \
   	y_3 = \frac{g}{F}, \\
   	z_3 &= - \,\frac{yF - xf}{F}
   \end{aligned}$
  } &
  \parbox{2cm}{
   $\psi_{\rm b} = \begin{bmatrix} F\\ f\\ g\end{bmatrix}$
  } &
  \parbox{6.5cm}{
  $\dsp L_{\rm b} = \begin{bmatrix}
          -x_1 & 1 & 0\\
          -y_1 & 0 & 1\\
          \ell_{31} & \alpha_2x - \alpha_1 - z & x - \alpha_2
         \end{bmatrix}$,} \\
 \addlinespace[1em]
 \multicolumn{3}{l}{where $\ell_{31} = -(\alpha_2x - \alpha_1 - z) x_1 + (\alpha_2 - x) y_1 - \nG(-p,-k)$.}\\
 \addlinespace[1em]
 \midrule
 \multicolumn{3}{l}{Gauge transformations for $L_{\rm a}$ and $L_{\rm b}$ are given by}\\
 \multicolumn{3}{l}{\parbox{12.5cm}{
   \begin{tabular}{p{5.5cm} p{3.5cm} p{3.5cm}}
    $\begin{aligned}
     L_{\rm b} &= \mc{G}_1 L_{\rm a} \mc{G}^{-1}, \psi_{\rm b} = \mc{G} \psi_{\rm a}, \\
     L_{\rm a} &= \bar{\mc{G}}_1 L_{\rm b} \bar{\mc{G}}^{-1},  \psi_{\rm a} = \bar{\mc{G}} \psi_{\rm b},
    \end{aligned}$
    &
    $\mc{G} = \begin{bmatrix} 1 & 0 & 0 \\ \frac{y}{x} & 0 & \frac{1}{x} \\ 0 & 1 & 0 \end{bmatrix} $
    & \hfill
    $\bar{\mc{G}} = \begin{bmatrix} 1 & 0 & 0 \\ 0 & 0 & 1 \\ -y & x & 1 \end{bmatrix}, $\\
    \multicolumn{3}{l}{where $\bar{\mc{G}} = \mc{G}^{-1}$.}
   \end{tabular}
  }
 } \\
 \addlinespace[0.8em]
 \bottomrule[1pt]
\end{longtable}
\vspace*{-4mm}
\noindent
%
%%%  Generalized Hietarinta C-3a System
% fixed lower cases 09/16/2019
\subsubsection{Generalized Hietarinta C-3 system}
\label{sec:gHC3GaugeEquivalence}
 The generalized Hietarinta C-3 system \cite{ref:Zhangetal2012},
 \begin{equation}
  \label{E:gHC3}
  \begin{aligned}
   &z y_1 + x_1 - x = 0, \;\; z y_2 + x_2 - x = 0, \;\;\text{ and } \\
   &\nG(-a,-b)x_{12} - yz_{12} + z\left(\frac{\nG(-q,-b) y_2 z_1 - \nG(-p, -b) y_1 z_2}{z_1 - z_2}\right) = 0,
  \end{aligned}
 \end{equation}
 has edge constraint $\dsp x_1 = x - z y_1$ leading to $\dsp x_3 = x - z y_3$.
 Here,
 \begin{equation}
  \label{E:gHC313cc}
  \begin{aligned}
   x_{13} &= \frac{x_3z_1 - x_1z_3}{z_1 - z_3}, \;\;
   y_{13} = \frac{x_1 - x_3}{z_1 - z_3}, \text{ and } \\
   z_{13} &= \frac{\nG(-a,-b)(x_3z_1 - x_1z_3) + z\Big(\nG(-k,-b)y_3z_1 - \nG(-p,-b)y_1z_3\Big)}{y(z_1 - z_3)}.
  \end{aligned}
 \end{equation}
 The $3$ by $3$ Lax matrices with their gauge transformations are given in Table 5.
 For $L_{\rm a}$ and $L_{\rm b}$ we set $t = s = \frac{1}{z}$
 since $\frac{s \, t_2}{t \, s_1} \; \dot{=}\; \frac{z_1}{z_2} $.
% fixed lower cases in Table caption 09/16/2019
% TABLE 5
\begin{longtable}{lcl}
% \caption{Gen. Hietarinta C-3 System Lax Pairs \& Gauge Matrices}
\caption{$\!$Generalized Hietarinta C-3 system Lax pairs and gauge matrices$\!$}
 \label{T:gHC3}\\
 \toprule[1pt]
  Substitutions & $\psi$ & Matrices $L$ of Lax pair \\
 \midrule
 \multicolumn{3}{l}{Writing the edge constraint as $x_3 = x - z y_3$ yields}\\
  \parbox{3.5cm}{
   $\begin{aligned}
   	x_3 &= \frac{xF - zg}{F}, \\
   	y_3 &= \frac{g}{F},\ \
   	z_3 = \frac{h}{F}.
   \end{aligned}$
  } &
  \parbox{2.5cm}{
   $\psi_{\rm a} = \begin{bmatrix} F\\ g\\ h\end{bmatrix}$
  } &
  \parbox{6.5cm}{ \hfill
  $\dsp L_{\rm a} = \frac{1}{z}\begin{bmatrix}
          -z_1 & 0 & 1\\
          x - x_1 & -z & 0\\
          -\nG(-a,-b)\frac{x z_1}{y} & \ell_{32} & \ell_{33}
         \end{bmatrix}$,} \\
 \addlinespace[0.5em]
 \multicolumn{3}{l}{\parbox{12.75cm}{with $\ell_{32} \!=\! \big( \nG(-a,-b) -\! \nG(-k,-b) \big)\frac{z z_1}{y},
  \ell_{33} \!=\! \big( \nG(-a,-b)x_1 +\! \nG(-p,-b) z y_1 \big)\frac{1}{y}$.}} \\
 \addlinespace[0.75em]
 \midrule
 \multicolumn{3}{l}{Writing the edge constraint as $y_3 = \frac{x - x_3}{z}$ yields}\\
  \parbox{3.5cm}{
   $\begin{aligned}
   	x_3 &= \frac{f}{F},\ \ z_3 = \frac{h}{F}, \\
   	y_3 &= \frac{xF - f}{zF}.
   \end{aligned}$
  } &
  \parbox{2.5cm}{
   $\psi_{\rm b} = \begin{bmatrix} F\\ f\\ h\end{bmatrix}$
  } &
  \parbox{6.5cm}{ \hfill
  $\dsp L_{\rm b} = \frac{1}{z}\begin{bmatrix}
          -z_1 & 0 & 1\\
          0 & -z_1 & x_1\\
          -\nG(-k,-b)\frac{x z_1}{y} & \tilde{\ell}_{32} & \tilde{\ell}_{33}
         \end{bmatrix}$,} \\
 \addlinespace[0.5em]
  \multicolumn{3}{l}{\parbox{12.85cm}{with $\tilde{\ell}_{32} \!=\! \big(\nG(-k,-b) -\! \nG(-a,-b)\big)\frac{z_1}{y},
   \tilde{\ell}_{33} \!=\! \big( \nG(-a, -b) x_1 + \nG(-p, -b) z y_1 \big)\frac{1}{y}$.}} \\
 \addlinespace[0.75em]
 \midrule
 \multicolumn{3}{l}{Gauge transformations for $L_{\rm a}$ and $L_{\rm b}$ are given by}\\
 \multicolumn{3}{l}{\parbox{12.5cm}{
   \begin{tabular}{p{5cm} p{3.5cm} p{4cm}}
    $\begin{aligned}
     L_{\rm b} &= \mc{G}_1 L_{\rm a} \mc{G}^{-1}, \psi_{\rm b} = \mc{G} \psi_{\rm a}, \\
     L_{\rm a} &= \bar{\mc{G}}_1 L_{\rm b} \bar{\mc{G}}^{-1},  \psi_{\rm a} = \bar{\mc{G}} \psi_{\rm b},
    \end{aligned}$
    &
    $\mc{G} = \begin{bmatrix} 1 & 0 & 0 \\ x & -z & 0 \\ 0 & 0 & 1 \end{bmatrix} $
    & \hfill
    $\bar{\mc{G}} = \begin{bmatrix} 1 & 0 & 0 \\ \frac{x}{z} & -\,\frac{1}{z} & 0 \\ 0 & 0 & 1 \end{bmatrix}. $\\
    \multicolumn{3}{l}{where $\bar{\mc{G}} = \mc{G}^{-1}$.}
   \end{tabular}
  }
 } \\
 \addlinespace[0.75em]
 \bottomrule[1pt]
\end{longtable}
\vspace*{-1mm}
\noindent
Incorporating edge constraints into \eqref{E:gHC313cc} yields
 \begin{equation}
  \label{E:gHC313sA}
  \begin{aligned}
   \tilde{x}_{13} &= x + \frac{z(y_1 z_3 - y_3 z_1)}{z_1 - z_3}, \;\;
       \tilde{y}_{13} = \frac{z(y_3 - y_1)}{z_1 - z_3}, \;\;\text{ and } \\
   \tilde{z}_{13} &= \frac{\nG(-a,-b)\big(x(z_1 - z_3) + z(y_1 z_3 - y_3 z_1)\big)}{y(z_1 - z_3)} \\
       &
         \;\;\;\; + \frac{z\Big(\nG(-k,-b) y_3 z_1 - \nG(-p,-b) y_1 z_3 \Big)}{y(z_1 - z_3)}.
  \end{aligned}
 \end{equation}
\vskip 1pt
\noindent
 System \eqref{E:gHC3} also admits extended Lax matrices:
 \begin{equation}
  \label{M:C3Lyz}
  L_{\rm A} = \frac{1}{z}
   \begin{bmatrix}
    -z_1 & 0 & 0 & 1\\
    0 & -z_1 & 0 & x_1\\
    zy_1 & 0 & -z & 0\\
    - \nG(-a,-b) \frac{xz_1}{y} & 0 & \big(\nG(-a,-b) - \nG(-k,-b)\big)\frac{zz_1}{y} & \ell_{44}
   \end{bmatrix},
 \end{equation}
 when considering the edge-modified solutions for $y_{13}$ and $z_{13}$ and where \\
 $\ell_{44} =  \left( \nG(-a,-b) x - \big( \nG(-a,-b) - \nG(-p,-b) \big) z y_1 \right) \frac{1}{y} $;
 and
 \begin{equation}
  \label{M:C3Lxyz}
  L_{\rm B} = \frac{1}{z}
   \begin{bmatrix}
    -z_1 & 0 & 0 & 1\\
    -x z_1 & 0 & z z_1 & x - z y_1 \\
    zy_1 & 0 & -z & 0\\
    -\nG(-a,-b)\frac{xz_1}{y} & 0 & \big(\nG(-a,-b) - \nG(-k,-b)\big)\frac{zz_1}{y} & \ell_{44}
   \end{bmatrix},
 \end{equation}
 when considering the edge-modified solutions \eqref{E:gHC313sA} and with $\ell_{44}$ as above;
 and
 \begin{equation}
  \label{M:C3Ly}
  L_{\rm C} = \frac{1}{z}
   \begin{bmatrix}
    -z_1 & 0 & 0 & 1\\
    0 & -z_1 & 0 & x_1\\
    zy_1 & 0 & -z & 0\\
    0 & - \nG(-a,-b)\frac{z_1}{y} & -\nG(-k,-b)\frac{zz_1}{y} & \tilde{\ell}_{44}
   \end{bmatrix},
 \end{equation}
 when considering the edge-modified solutions for $y_{13}$ and where
$\tilde{\ell}_{44} \!=\! \left( \nG(-a,-b)x_1 \right.$
\newline
$ \left. + \nG(-p,-b) z y_1 \right)\frac{1}{y} $.
The matrix $L_{\rm C}$ was derived in \cite[eq.\ (95)]{ref:Zhangetal2012} using $\tilde{y}_{13}$.
 All other combinations of \eqref{E:gHC313cc} and \eqref{E:gHC313sA} result in matrices which do not satisfy
 the defining equation \eqref{matLaxEqdiffeqs}.
%
%%%  Generalized Hietarinta C-4 System
% fixed lower cases 09/16/2019
\subsubsection{Generalized Hietarinta C-4 system}
\label{sec:HC4GaugeEquivalence}
 The generalized Hietarinta C-4 system \cite{ref:Zhangetal2012} is given by
 \begin{equation}
  \label{E:HC4}
  \begin{aligned}
   &z y_1 + x_1 - x = 0, \;\;
       z y_2 + x_2 - x = 0, \;\text{ and } \\
   &y z_{12} - z\left(\frac{\mG(p) y_1 z_2 - \mG(q) y_2 z_1}{z_1 - z_2}\right)
              - x x_{12} + \frac{1}{4}\nG(-a,-b)^2 = 0,
  \end{aligned}
 \end{equation}
 where
 \begin{equation}
\label{E:calGHC4}
 \dsp \mG(\tau) := -\,\frac{1}{2}\big(\nG(-\tau,-a) + \nG(-\tau,-b)\big).
 \end{equation}
 Edge constraint $\dsp x_1 = x - z y_1$ yields $\dsp x_3 = x - z y_3$.
 Here,
 \begin{equation}
  \label{E:gHC413cc}
  \begin{aligned}
   x_{13} &= \frac{x_3z_1 - x_1z_3}{z_1 - z_3}, \;\; y_{13} = \frac{x_1 - x_3}{z_1 - z_3}, \;\; \text{ and }\\
   z_{13} &= \frac{x(x_3z_1 - x_1z_3) - z\big(\mG(k) y_3 z_1 - \mG(p) y_1 z_3\big) }{y(z_1 - z_3)}
      - \frac{\nG(-a,-b)^2}{4y}.
  \end{aligned}
 \end{equation}
\noindent
 Variants of \eqref{E:gHC413cc} obtained by incorporating edge constraints are
 \begin{equation}
  \label{E:gHC413sA}
  \begin{aligned}
   \tilde{x}_{13} &= \frac{x(z_1 - z_3) - z(y_3 z_1 - y_1 z_3)}{z_1 - z_3}, \;\;
     \tilde{y}_{13} = -\, \frac{z(y_1 - y_3)}{z_1 - z_3}, \;\; \text{ and } \\
   \tilde{z}_{13} &= \frac{x z (y_1 z_3 - y_3 z_1) - z\big(\mG(k) y_3 z_1 - \mG(p) y_1 z_3\big) }{y(z_1 - z_3)}
             + \frac{4x^2 - \nG(-a,-b)^2}{4y}.
  \end{aligned}
 \end{equation}
 The $3$ by $3$ Lax matrices and the gauge transformations are given in Table 6.
 For $L_{\rm a}$ and $L_{\rm b}$ we take $t = s = \frac{1}{z}$
 since $\frac{s \, t_2}{t \, s_1} \; \dot{=}\; \frac{z_1}{z_2} $.
% fixed lower cases in Table caption 09/16/2019
% TABLE 6
\begin{longtable}{lcl}
% \caption{Gen. Hietarinta C-4 System Lax Pairs \& Gauge Matrices}
\caption{$\!$Generalized Hietarinta C-4 system Lax pairs and gauge matrices$\!$}
 \label{T:gHC4}\\
 \toprule[1pt]
  Substitutions & $\psi$ & Matrices $L$ of Lax pair \\
 \midrule
 \multicolumn{3}{l}{Writing the edge constraint as $x_3 = x - zy_3$ yields}\\
 \addlinespace[1em]
  \parbox{3.5cm}{
   $\begin{aligned}
   	x_3 &= \frac{xF - zg}{F}, \\
   	y_3 &= \frac{g}{F}, \hspace{3mm}
   	z_3 = \frac{h}{F}.
   \end{aligned}$
  } &
  \parbox{2.5cm}{
   $\psi_{\rm a} = \begin{bmatrix} F\\ g\\ h\end{bmatrix}$
  } &
  \parbox{6.5cm}{
  $\dsp L_{\rm a} = \frac{1}{z}\begin{bmatrix}
          -z_1 & 0 & 1\\
          x - x_1 & -z & 0\\
          \ell_{31}  & \frac{z z_1}{y}\left(x + \mG(k)\right) & \ell_{33}
         \end{bmatrix}$,} \\
 \addlinespace[1em]
 \multicolumn{3}{l}{
\parbox{13.25cm}{with $\ell_{31} \!=\! -\frac{z_1}{y}\big(x^2 - \frac{1}{4}\nG(-a,-b)^2\big),\,
\ell_{33} \!=\!\frac{1}{y} \big( x x_1 -\! \frac{1}{4}\nG(-a,-b)^2 -\! \mG(p) z y_1 \big)$.}
  }\\
 \addlinespace[0.75em]
 \midrule
 \multicolumn{3}{l}{Writing the edge constraint as $y_3 = \frac{x - x_3}{z}$ yields}\\
  \parbox{3.5cm}{
   $\begin{aligned}
   	x_3 &= \frac{f}{F},\hspace{3mm} z_3 = \frac{h}{F}\\
   	y_3 &= \frac{xF - f}{zF}. \\
   \end{aligned}$
  } &
  \parbox{2.5cm}{
   $\psi_{\rm b} = \begin{bmatrix} F\\ f\\ h\end{bmatrix}$
  } & \hfill
  \parbox{6.5cm}{ \hfill
  $\dsp L_{\rm b} = \frac{1}{z}\begin{bmatrix}
          -z_1 & 0 & 1\\
          0 & -z_1 & x_1\\
          \tilde{\ell}_{31} & - \frac{z_1}{y}\left(x + \mG(k)\right) & \tilde{\ell}_{33}
         \end{bmatrix}$.} \\
 \addlinespace[0.75em]
 \multicolumn{3}{l}{
  \parbox{13.5cm}{with $\tilde{\ell}_{31} \!=\! \frac{z_1}{y}\big( \frac{1}{4}\nG(-a,-b)^2 +\! \mG(k) x \big),
  \tilde{\ell}_{33} \!=\! \frac{1}{y}\big(x x_1 -\! \frac{1}{4} \nG(-a,-b)^2 -\! \mG(p) z y_1 \big)$.}
  }\\
 \addlinespace[0.75em]
 \midrule
 \multicolumn{3}{l}{Gauge transformations for $L_{\rm a}$ and $L_{\rm b}$ are given by}\\
 \multicolumn{3}{l}{\parbox{12.5cm}{
   \begin{tabular}{p{5cm} p{3.5cm} p{4cm}}
    $\begin{aligned}
     L_{\rm b} &= \mc{G}_1 L_{\rm a} \mc{G}^{-1}, \psi_{\rm b} = \mc{G} \psi_{\rm a}, \\
     L_{\rm a} &= \bar{\mc{G}}_1 L_{\rm b} \bar{\mc{G}}^{-1},  \psi_{\rm a} = \bar{\mc{G}} \psi_{\rm b},
    \end{aligned}$
    &
    $ \mc{G} = \begin{bmatrix} 1 & 0 & 0 \\ x & -z & 0 \\ 0 & 0 & 1 \end{bmatrix} $
    & \hfill
    $ \bar{\mc{G}} = \begin{bmatrix} 1 & 0 & 0 \\ \frac{x}{z} & -\,\frac{1}{z} & 0 \\ 0 & 0 & 1 \end{bmatrix}.$ \\
    \multicolumn{3}{l}{where $\bar{\mc{G}} = \mc{G}^{-1}$.}
   \end{tabular}
  }
 } \\
 \addlinespace[0.50em]
 \bottomrule[1pt]
\end{longtable}
\vspace*{3mm}
\noindent
System \eqref{E:HC4} has the following extended Lax matrices:
\begin{equation}
  \label{M:C4Lyz}
  L_{\rm A} = \frac{1}{z}
   \begin{bmatrix}
   -z_1 & 0 & 0 & 1\\
   0 & -z_1 & 0 & x_1\\
   zy_1 & 0 & -z & 0\\
   \ell_{41} & 0 & \frac{z z_1}{y}\big(x + \mG(k)\big) & \ell_{44}
  \end{bmatrix},
 \end{equation}
 when considering edge-modified $y_{13}$ and $z_{13},$ and where
 $\ell_{41} \!=\! -\frac{z_1}{y}\big(x^2 - \frac{1}{4}\nG(-a,-b)^2\big)$
 and
 $\ell_{44} = \frac{1}{y}\big( x^2 - \frac{1}{4}\nG(-a,-b)^2
              - z y_1(x + \mG(p))\big)$;
 \begin{equation}
  \label{M:C4Lxyz}
   L_{\rm B} = \frac{1}{z}
    \begin{bmatrix}
     - z_1 & 0 & 0 & 1\\
     - xz_1 & 0 & zz_1 & x - zy_1 \\
     zy_1 & 0 & -z & 0\\
     \ell_{41} & 0 & \frac{z z_1}{y}\big(x + \mG(k)\big) & \ell_{44}
    \end{bmatrix},
 \end{equation}
 by taking the edge-modified expression of \eqref{E:gHC413sA}, with $\ell_{41}$ and $\ell_{44}$
 as above; and
 \begin{equation}
  \label{M:C4Ly}
  L_{\rm C} = \frac{1}{z}
   \begin{bmatrix}
    -z_1 & 0 & 0 & 1\\
    0 & -z_1 & 0 & x_1\\
    zy_1 & 0 & -z & 0\\
    \frac{z_1 \nG(-a,-b)^2}{4y} & -\frac{x z_1}{y} & \frac{z z_1}{y} \mG(k) & \tilde{\ell}_{44}
   \end{bmatrix}
 \end{equation}
 when using the edge-modified expression for $y_{13}$ and with
$\tilde{\ell}_{44} \!=\! \frac{1}{y} \big( x x_1 -\frac{1}{4}\nG(-a,-b)^2 - \mG(p) z y_1 \big) $.
All other combinations of \eqref{E:gHC413cc} and \eqref{E:gHC413sA} result in matrices
that fail to satisfy the defining equation \eqref{matLaxEqdiffeqs}.
%
% fixed lower cases 09/16/2019
% SECTION 7 -- DISCUSSION AND CONCLUSIONS
\section{Software implementation and conclusions}
\label{sec:DiscussionConclusions}
The method to find Lax pairs of \PDeltaEs\ based on multi-dimensional consistency
is being implemented in {\sc Mathematica}.
Using our prototype {\sc Mathematica} package \cite{ref:Bridgman2015} we derived Lax matrices of
minimal sizes for various Boussinesq-type equations.
% fixed chapter instead of paper 09/16/2019
In turn, the research done for this chapter helped us improve and extend the capabilities of the
software under development \cite{ref:BridgmanHereman2019}.

The way we symbolically compute (and verify) Lax pairs might slightly differ from the
procedure used by other authors (by hand or interactively with a computer algebra system).
Indeed, for a system of \PDeltaEs, the software generates all equations (and solutions)
necessary to define a full face of the quadrilateral.
That is, for a system of \PDeltaEs\ including full-face expressions
(involving at least 3 corners of the quadrilateral) and edge equations
(involving two adjacent corners of the quadrilateral),
the software will first augment the given system with the additional edge equations necessary
to complete the set of equations for a particular face of the cube.
For example, the Schwarzian Boussinesq system \eqref{E:sBouss} discussed in Section~\ref{E:sBouss}
is augmented with two additional edge equations,
% fixed comma instead of period 09/16/2019
\begin{equation}
\label{E:a2HietFrontExtraEdges}
z_2 y_{12} - x_{12} + x_2 = 0, \;\;
z_1 y_{12} - x_{12} + x_1 = 0,
\end{equation}
to generate the full set of equations for the front face of the cube.
Then, using lexicographical ordering ($x \prec y \prec z$) and an index ordering
(double-subscripts $\prec$ no-subscripts $\prec$ single-subscripts), the
software solves
\eqref{E:sBouss} and \eqref{E:a2HietFrontExtraEdges} yielding
\begin{subequations}
 \label{E:sBoussfrontfacesols}
 \begin{align}
 x_{12} &= \frac{x_2 z_1 - x_1 z_2}{z_1 - z_2}, \;\; y_{12} = \frac{x_2 - x_1}{z_1 - z_2}, \;\;
 z_{12} = \frac{z}{y} \left( \frac{p y_1z_2 - q y_2z_1}{z_1 - z_2} \right), \; \text{ with }
 \label{E:sBoussfrontfacesolsx12y12} \\
 x &= x_1 - z y_1, \; \text{ and } \; z = \frac{x_1 - x_2}{y_1 - y_2}.
 \label{E:sBoussfrontfacesolsz12xy}
 \end{align}
\end{subequations}
This process is then repeated for the left and bottom faces of the cube, always substituting
solutions such as \eqref{E:sBoussfrontfacesols} to enforce consistency and remove redundancies.

The complete front-corner system (i.e., front, left and bottom faces connected at corner ${\bf x}$,
see Fig.~\ref{fig:cube}) is used to simplify the Lax equation when a Lax pair is finally tested.
If all works as planned, the evaluation of the Lax equation then automatically results in a zero matrix.

For verification of consistency about the cube, the equations for the faces connected at the back-corner
(where ${\bf x}_{123}$ is located as shown in Fig.~\ref{fig:cube}) are computed and then solved
(adhering to the above ordering).
Next, these solutions are then checked for consistency with the front-corner system.
Finally, if the system is 3D consistent, the multiple expressions obtained for ${\bf x}_{123}$
should be equal when reduced using \eqref{E:sBoussfrontfacesols} augmented with like equations for
${\bf x}_{13}$ and ${\bf x}_{23}.$

Why would one care about different Lax matrices, in particular, if they are gauge equivalent?
In the PDE case, application of the IST is easier if one selects a Lax pair of a specific form
(i.e., the eigenvalues should appear in the diagonal entries), chosen from the infinite number
of gauge equivalent pairs.
Thus, for the KdV equation one may prefer to work with \eqref{E:kdvLAX2} instead of \eqref{E:kdvLAX1}.
Similar issues arise for \PDeltaEs.
Among the family of gauge-equivalent Lax matrices for \PDeltaEs, which one should be selected
so that, for example, the IST or staircase method \cite{ref:vanderKampQuispel2010} could be applied?
(The latter method is used to find first integrals for periodic reductions of integrable \PDeltaEs).
In addition, one has to select an appropriate (separation) factor $t({\bf x}, {\bf x}_1; p, k)$
(see Sec.~\ref{sec:firstAlt1}).
% fixed chapter instead of paper 09/16/2019
These issues are not addressed in this chapter for they require further study.
% fixed Acknowledgments instead of Acknowledgements 09/16/2019
\vspace*{-3mm}
\section*{Acknowledgments}
\label{sec:Acknowledgements}
This material is based in part upon research supported by the National Science Foundation (NSF)
under Grant No.\ CCF-0830783.
% fixed added the 09/16/2019
Any opinions, findings, and conclusions or recommendations expressed in this material are those
of the authors and do not necessarily reflect the views of the NSF.


\vspace*{-3mm}
\begin{thebibliography}{99}

% verified
\bibitem{ref:AblowitzClarkson1991}
Ablowitz M J and Clarkson P A,
{\it Solitons, Nonlinear Evolution Equations and Inverse Scattering},
(London Math. Soc. Lect. Note Ser. {\bf 149})
Cambridge Univ.\ Press, Cambridge, UK, 1991.

% verified
\bibitem{ref:AKNS1974}
Ablowitz M J, Kaup D J, Newell A C, and Segur H,
The inverse scattering transform--Fourier analysis for nonlinear problems,
{\it Stud. Appl. Math.} {\bf 53}(4), 249--315, 1974.

% verified
\bibitem{ref:AblowitzLadik1976}
Ablowitz M J and Ladik F J,
A nonlinear difference scheme and inverse scattering,
{\it Stud. Appl. Math.} {\bf 55}(3), 213--229, 1976.

% verified
\bibitem{ref:AblowitzLadik1977}
Ablowitz M J and Ladik F J,
On the solution of a class of nonlinear partial difference equations,
{\it Stud. Appl. Math.} {\bf 57}(1), 1--12, 1977.

% verified
\bibitem{ref:Adler2003}
Adler V E, Bobenko A I, and Suris Yu B,
Classification of integrable equations on quad-graphs. The consistency approach,
{\it Commun. Math. Phys.} {\bf 233}(3), 513--543, 2003.

% verified
\bibitem{ref:BobenkoSuris2002}
Bobenko A I and Suris Yu B,
Integrable systems on quad-graphs,
{\it Int. Math. Res. Not.} {\bf 2002}(11), 573--611, 2002.

% verified
\bibitem{ref:BobenkoSuris2008}
Bobenko A I and Suris Yu B,
{\it Discrete Differential Geometry: Integrable Structure},
(Grad. Stud. Math. {\bf 98}) AMS, Philadelphia, PA, 2008.

% partially fixed  left code name the same but removed sc on code name and Mathematica 09/16/2019
% verified
\bibitem{ref:Bridgman2015}
Bridgman T J,
% {\sc LaxPairPartialDifferenceEquations.m}:
\verb|LaxPairPartialDifferenceEquations.m|:
a Mathematica package for the symbolic computation
of Lax pairs of systems of nonlinear partial difference equations defined on quadrilaterals,
\phantom{xxxxxxxxxxxxxxxxx}
\verb|http://inside.mines.edu/~whereman/software/LaxPairPartialDifference|
\newline
\noindent
\verb|Equations|, 2012-2019.

% verified
\bibitem{ref:Bridgman2018}
Bridgman T J,
{\it Symbolic Computation of Lax Pairs of Nonlinear Partial Difference Equations},
Ph.D Thesis, Dept. Appl. Maths. Stats., Colorado School of Mines, 2018.

% verified
\bibitem{ref:BridgmanHereman2018}
Bridgman T and Hereman W,
Gauge-equivalent Lax pairs for Boussinesq-type systems of partial difference equations,
Report, Dept. Appl. Maths. Stats., Colorado School of Mines, 26 pages, submitted, 2018.

% verified
\bibitem{ref:BridgmanHereman2019}
Bridgman T and Hereman W,
Symbolic software for the computation of Lax pairs of nonlinear partial difference equations,
in preparation, 2019.

% verified
\bibitem{ref:Bridgmanetal2013}
Bridgman T, Hereman W, Quispel G R W, and van der Kamp P H,
Symbolic computation of Lax pairs of partial difference equations using consistency around the cube,
{\it Found. Comput. Math.} {\bf 13}(4), 517--544, 2013.

% verified
\bibitem{ref:FaddeevTakhtajan1987}
Faddeev L D and Takhtajan L A,
{\it Hamiltonian Methods in the Theory of Solitons},
(Springer Ser. Sov. Math.) Springer-Verlag, Berlin, Germany, 1987.

% verified
\bibitem{ref:Hickmanetal2012}
Hickman M, Hereman W, Larue J, and G\"{o}kta\c{s} \"{U},
Scaling invariant Lax pairs of nonlinear evolution equations,
{\it Applicable Analysis} {\bf 91}(2), 381--402, 2012.

% verified
\bibitem{ref:Hietarinta2011}
Hietarinta J,
Boussinesq-like multi-component lattice equations and multi-dimensional consistency,
{\it J. Phys. A: Math. Theor.} {\bf 44}(16), 165204, 22 pages, 2011.

% verified
\bibitem{ref:Hietarinta2018}
Hietarinta J,
Elementary introduction to discrete soliton equations.
In: Euler N (Ed),
{\it Nonlinear Systems and Their Remarkable Mathematical Structures} {\bf 1},
Chapter A4, 74--93, CRC Press, Boca Raton, Florida, 2018.

% verified
\bibitem{ref:HietarintaJoshiNijhoff2016}
Hietarinta J, Joshi N, and Nijhoff F W,
{\it Discrete Systems and Integrability},
(Cambridge Texts Appl. Math.) Cambridge Univ. Press, Cambridge, UK, 2016.

% verified
\bibitem{ref:HirotaI}
Hirota R,
Nonlinear partial difference equations. I - A difference analogue of the Korteweg-de Vries equation,
{\it J. Phys. Soc. Jpn.} {\bf 43}(4), 1424--1433, 1977.

% verified
\bibitem{ref:Hirota2004}
Hirota R,
{\it The Direct Methods in Soliton Theory},
(Cambridge Tracts Math.)
Cambridge Univ. Press, Cambridge, UK, 2004.

% verified
\bibitem{ref:Hydon2014}
Hydon P E,
{\it Difference Equations by Differential Equation Methods},
(Cambridge Monographs Appl. Comp. Math. {\bf 27})
Cambridge Univ. Press, Cambridge, UK, 2014.

% verified
\bibitem{ref:Levietal2011}
Levi D, Olver P, Thomova Z, and Winternitz P (Eds),
{\it Symmetries and Integrability of Difference Equations},
(London Math. Soc. Lect. Note Ser. {\bf 381})
Cambridge Univ. Press, Cambridge, UK, 2011.

% verified
\bibitem{ref:Levietal2017}
Levi D, Verge-Rebelo R, and Winternitz P (Eds),
{\it Symmetries and Integrability of Difference Equations},
% (Lect. Notes Abecederian School of SIDE {\bf 12}, Montreal 2016,
(CRM Ser.\ Math. Phys.) Springer Int. Publ., New York, 2017.

% verified
\bibitem{ref:Miura1968}
Miura R M,
Korteweg-de Vries equation and generalizations. I. A remarkable explicit nonlinear transformation,
{\it J. Math. Phys.} {\bf 9}(8), 1202--1204, 1968.

% verified
\bibitem{ref:Nijhoff1997}
Nijhoff F W,
On some ``Schwarzian'' equations and their discrete analogues.
In: Fokas A S and Gelfand I M (Eds),
{\it Algebraic Aspects of Integrable Systems: In memory of Irene Dorfman}, 237--260,
Birkh{\"a}user Verlag, Boston, MA, 1997.

% verified
\bibitem{ref:Nijhoff1999}
Nijhoff F W,
Discrete Painlev\'e equations and symmetry reduction on the lattice.
In: Bobenko A I and Seiler R (Eds),
{\it Discrete Integrable Geometry and Physics}, 209--234,
(Oxford Lect. Ser. Math. Appls.) Oxford Univ. Press, New York, 1999.

% verified
\bibitem{ref:Nijhoff2002}
Nijhoff F W,
Lax pair for the Adler (lattice Krichever-Novikov) system,
{\it Phys. Lett. A} {\bf 297}(1-2), 49--58, 2002.

% verified
\bibitem{ref:NijhoffCapel1995}
Nijhoff F and Capel H,
The discrete Korteweg-de Vries equation,
{\it Acta Appl. Math.} {\bf 39}(1-3), 133--158, 1995.

% verified
\bibitem{ref:Nijhoff1991}
Nijhoff F W and Papageorgiou V G,
Similarity reductions of integrable lattices and discrete analogues of the Painlev\'e II equation,
{\it Phys. Lett. A} {\bf 153}(6-7), 337--344, 1991.

% verified
\bibitem{ref:Nijhoffetal1992}
Nijhoff F W, Papageorgiou V G, Capel H W, and Quispel G R W,
The lattice Gel'fand-Dikii hierarchy,
{\it Inv. Probl.} {\bf 8}(4), 597--621, 1992.

% verified -- this one is correct!
\bibitem{ref:Nijhoff1983}
Nijhoff F W, Quispel G R W, and Capel H W,
Direct linearization of nonlinear difference-difference equations,
{\it Phys. Lett. A} {\bf 97}(4), 125--128, 1983.

% verified
\bibitem{ref:Nijhoff1982}
Nijhoff F W, van der Linden J, Quispel G R W, Capel H W, and Velthuizen J,
Linearization of the nonlinear Schr\"{o}dinger equation and the isotropic Heisenberg spin chain,
{\it Physica A} {\bf 116}(1-2), 1--33, 1982.

% fixed lower case 09/16/2019
% verified
\bibitem{ref:Tongas2005}
Tongas A and Nijhoff F W,
The Boussinesq integrable system: compatible lattice and continuum structures,
{\it Glasgow Math. J.} {\bf 47}(A), 205--219, 2005.

% verified
\bibitem{ref:vanderKampQuispel2010}
van der Kamp P H and Quispel G R W,
The staircase method: integrals for periodic reductions of integrable lattice equations,
{\it J.\ Phys. A: Math. Theor.} {\bf 43}(46), 465207, 34 pages, 2010.

% verified
\bibitem{ref:Wahlquist1973}
Wahlquist H D and Estabrook F B,
B\"acklund transformation for solutions of the Korteweg-de Vries equation,
{\it Phys. Rev. Lett.} {\bf 31}(23), 1386--1390, 1973.

% verified
\bibitem{ref:Zhangetal2012}
Zhang D-J, Zhao S-L, and Nijhoff F W,
Direct linearization of extended lattice BSQ systems,
{\it Stud. Appl. Math.} {\bf 129}(2), 220--248, 2012.

\end{thebibliography}
\end{document}